\documentclass[
superscriptaddress,
pra,
showpacs,
showkeys,
aps,
floatfix,
]{revtex4-1}

\usepackage{amsmath}
\usepackage{amssymb}

\usepackage{graphicx}
\usepackage{bm,bbm}
\usepackage{color}
\usepackage{placeins}
\usepackage[normalem]{ulem}
\usepackage{soul}

\newcommand{\eg}{\textit{e.g.}}
\newcommand{\ie}{\textit{i.e.} }

\newcommand{\mt}{\mathrm }

\begin{document}

\title{Stochastic epidemiological model: Modeling the SARS-CoV-2
spreading in Mexico}

\author{P. C. L\' opez V\' azquez}
\affiliation{Departamento de Ciencias Naturales y Exactas, Universidad de Guadalajara,
Carretera Guadalajara - Ameca Km. 45.5 C.P. 46600. Ameca, Jalisco, M\'exico.}

\author{G. S\'anchez-Gonz\'alez}
\affiliation{Centro de Investigaci\'on Sobre Enfermedades Infecciosas,
Instituto Nacional de Salud P\'ublica, Universidad No. 655,
 C.P. 62100, Cuernavaca, Morelos. M\'exico.}

\author{J. Mart\'inez-Ortega}
\affiliation{Coordinación General de Innovación Gubernamental,
  Gobierno del Estado de Jalisco , Ciudad Creativa Digital, Independencia No. 55
  Piso 5, Col. Centro, Guadalajara, Jal. CP. 44100}

\author{R. S. Arroyo-Duarte}
\affiliation{Coordinación de Análisis Estratégico,
Gobierno del Estado de Jalisco , Ciudad Creativa Digital, Independencia No. 55
Piso 5, Col. Centro, Guadalajara, Jal. CP. 44100}


\begin{abstract}
In this paper we model the spreading of the SARS-CoV-2 in Mexico by
introducing a new stochastic approximation
constructed from first principles, structured on the basis of a
Latent-Infectious-(Recovered or Deceased) (LI(RD)) compartmental
approximation, where the number of new infected individuals caused by a
single infectious individual per unit time (a day), is a random variable
of a Poisson distribution and whose parameter is modulated through a
weight-like time-dependent function. The weight function serves to
introduce a time dependence to the average number of new infections and as
we will show, this information can be extracted from empirical data,
giving to the model self-consistency and provides a tool to study
information about periodic patterns encoded in the epidemiological
dynamics.
\end{abstract}
 \maketitle

 \section*{Introduction}
 Since the late 2019 to the date, the rapid worldwide spread of the SARS-CoV-2
 has caused around four and a half million of human deaths ~\cite{WHO1}, placing
 mankind in one of the most challenging episodes in the recent human history.
 An extraordinary effort has been made to implement mathematical methods that
 could give accurate approximations to the spreading of the epidemia, looking to
 forecast and to implement non-pharmaceutical responses to reduce the damage in
 the society~\cite{Shan21}. These methods ranging from standard compartmental
 models (typically employed at the beginning of the epidemia to determine the
 epidemiological parameters~\cite{Hy13,Re21,Ca20, Ta20, Da21, Wu20}), to hybrid
 methods that incorporates stochastic meta-population network models with local
 and global mobility patterns~\cite{Ri20,Sa20,Ch20,Wo20,Ch21}; attempt to
 overcome the complex behavior of social interaction characterized by the
 tendency of the population to cluster~\cite{Borr17}, following quasi-periodic
 patterns of mobility in large dense urbanized areas~\cite{Fu06,Wo20,Ch21}.

 In addition to the complexity for determining the degree of connectivity
 (the contact network) among individuals in urbanized areas, regulatory measures,
 such as home lockdowns and social distancing, (which provide an additional
 degree of complexity in determining the spreading of the disease), were promoted
 to reduce the transmission of the infection, in order to keep health services
 unsaturated~\cite{WHO2}.


 How and when to promote regulatory measures became some of the most difficult
 decisions followed by the different populations along the world because of their
 effects over health, economics and many other social factors.
 These decisions must be supported on predictive models possessing a good
 equilibrium between registered data (reliable readouts about registered
 confirmed cases), mobility patterns followed by the population, and
 computational efficiency of the epidemiological
 models~\cite{Ch20,Wu20,Gi20,Pu20,Ch21,Ba09,Da09}.
 Nevertheless, their efficiency relies on a good accessibility and
 characterization of the available data~\cite{Ch21}, which in the case of less
 developed countries, these data could be more difficult or impossible to obtain.
 In this regard, stochastic models, which introduce a randomization about certain
 unknown elements could provide an alternative guidance.

 In this paper, we introduce a new stochastic model which has served us to
 simulate and follow the spreading of the Sars-CoV-2 in Mexico.
 Unlike recent approximations employing stochastic models to
 to model the Sars-CoV-2 spreading, which are based on introducing additive
 white Gaussian noise to the contact parameter $\beta$ in the standard SIR model
 \cite{Dor21,Deb21}, or by considering a master equation following transition
 probabilities associated to the law of mass interaction governing the
 standard SIR model dynamics~\cite{Eng20}; here we propose a model which consist on
 deriving the dynamical equations of a LI(RD)
 compartmental model (Latent-Infectious-(Recovered or Deceased)),
 by employing a randomization about the number of infections caused by a
 single infected individual, per unit time (a day), together with a modulation of
 the  daily mean of infected population through a weight-like time dependent
 function. This modulation serves to
 introduce variations in the probability of infection caused by several
 phenomenological or fundamental behavior such as pharmaceutical or
 non-pharmaceutical interventions and herd immunity as well. In addition
 we will establish the relation that the weight function has with the effective
 reproduction number $\mathcal{R}(t)$ and derive a method to construct
 the weight function from empirical data, giving in this way self-consistency to
 the model. In other words, the model attempts to describe a scenario
 about how many people can infect one infectious individual per day when the
 infectious events are considered to be homogeneously distributed in time and
 when the probability of infection is affected by pharmaceutical or
 non-pharmaceutical interventions.

 Finally, through this model we analyze the evolution of the disease
 in some Mexican states, some of them housing the largest metropolitan areas of
 México.

 The paper is organized as follows: In the first section \ref{model}, we derive
 the stochastic compartmental model and introduce the weight function as a tool
 to modulate the mean of the daily infected individuals and show how
 the weight function can help us to incorporate certain effects emerging
 as a consequence of variations in the probability of
 infection such as herd immunity and confinement. Furthermore, we establish the
 connection between the weight function and the effective time-dependent
 reproduction number and derive an empirical estimation about the effective
 reproduction number. In section \ref{fcast} we employ our
 model to study the development of the COVID-19 in México.
 Finally in section \ref{con} we presents our conclusions.


 \section{\label{model} The model}

 The epidemiological model we propose, consist on the randomization of the number
 of infections caused daily. We use a time-dependent Poisson processes to
 generate the new infections caused by each of the infectious individuals,
 along a given period of time (the time unit); when restrictions on
 mobility are continuously changing in time.
 The core of the model is constructed on the basis of a compartmental
 description constituted by a susceptible population $S(t)$ which serves
 merely to have finite resource about where to choose randomly the number of new
 infections; the infected-latent population $L(t)$ which is randomly obtained
 and the infectious population $I(t)$ which represents the the part of the
 population capable of infecting.
 The employment of a Poisson process responds to the assumption that the new
 infectious events can be interpreted as homogeneously distributed infectious
 events in time.

 Once the number of infections per infected individual at a single time step
 (\ie, the daily infected (but not infectious) population per infected
 individual) has been obtained, they are removed from the susceptible condition
 an placed into the latent condition $L(t)$ which characterizes the part of the
 population that is infected but is not capable to transmit the virus until an
 incubation time (latency time $t_L$), has passed. After the latency time, the
 infected-latent population becomes contagious, passing into the infectious
 condition $I(t)$ associating to each member of this new group of infectious
 population, an individual contagious domain, hence becoming able to transmit
 the disease to the susceptible population.
 The number of the latent and the infectious population at time $t_{j+1}$ may be
 written as follows:
 \begin{eqnarray}\label{nl}
 L(t_{j+1}) &=& L(t_{j}) + \sum_{i=1}^{I(t_{j})} \{\chi(t_j)\}_i
  - \theta(t_{j}-t_L)\,L(t_{j}-t_L)\,,\\\nonumber
 &&\\\label{ni}
 I(t_{j+1}) &=& I(t_{j}) + \theta(t_{j}-t_L)\,L(t_{j}-t_L)\,,
 \end{eqnarray}
 where  $\{\chi(t_j)\}_i \leftarrow P(\lambda(t_j))$ is a random variable
 giving the amount of new infected individuals due to the $i$-th infectious
 individual at time $t_j$ (which with absolute certainty become infected),
 while the intensities of the Poisson process, (\ie~the parameters
 $\lambda(t_j)$), describe the mean number of contagious events
 at the time $t_j$. Additionally, in
 Eqs. (\ref{nl}, \ref{ni}), we have introduced the Heaviside function
 $\theta(\cdot)$ to start counting individuals after the latency time has passed.

 In a real scenario, the spreading of a disease depends on the degree of
 close contact among its individuals,
 which in turn depends on the degree of urbanization and mobility of the
 population~\cite{Ha21}, however, we believe that part of these complex
 aspects could be captured into our model by a proper parametrization of
 the Poisson processes, \ie, the daily mean number of infections per
 infectious individual $\lambda(t_j)$.
 In this regard, we parametrize the mean
 number of the daily infections by introducing a time-dependent function
 which indirectly serves to modify the daily probability of infection.

 In other words, to each time $t$ we associate the following mean of the number
 of infections produced daily:
 \begin{equation}\label{means}
   \lambda(t) = \varrho_o\, W(t)\,,
 \end{equation}
 where the parameter $\varrho_o$
 represents an initial estimations about the average of the number of infections
 that a single infectious individual can cause per unit time, \ie, the ratio
 between the basic reproduction number (calculated at the beginning of an
 epidemiological event~\cite{Li20,Gu20,Ca20}) and the infectious period:
 \begin{equation}
 \varrho_o = R_o /t_I\,,
 \end{equation}
 with the infectious period being described as $t_{I} = t_R-t_L$,
 with $t_R$ being the recovery time since the individual
 became firstly infected (although not contagious) and $t_L$ is the latency time.
 On the other hand, $W(t)$ appearing in (\ref{means}) represents a
 time-dependent weight function which serves to modulated the mean of the
 number of infections, \ie, $W(t)$ represents the daily variability of the
 probability of infection through the daily mean of new infections.
 We will address the employment of $W(t)$ in the following section.

 Finally, and following within the compartment direction, we consider that the
 infectious population could pass, either to the recovered $R(t)$ or to the
 deceased $D(t)$ condition, depending on the development of the disease in the
 infected individual. In the former, we define the recovery time $t_R$, after
 which the infected population heals. In this sense, the number of recovered
 population at the time $t_{j+1}$ is given by:
 \begin{equation}\label{nr}
 R(t_{j+1})  = R(t_{j}) + \theta(t_j-t_R)\,I(t_j-t_R)\,.
 \end{equation}

 For modeling the deceased population, we make use of an additional random
 procedure to select from each of the infected-latent individuals at the time
 $t_j$ and according to a given  fatality rate, the infected population that will
 pass to the deceased group; in other words, we count every new set of
 infected-latent individuals appearing at time $t_{j+1}$, \ie
 \begin{equation}
 L(t_{j+1}) - L(t_{j}) = \sum_{i=1}^{I(t_{j})} \chi_i(t_{j})\,,
 \end{equation}
 and for each of the new cases, we use a uniform distribution to generate a
 random number $r\in\mt{unif}(0,1)$ which is compared to $p=1-l$ where $l$ is
 the lethality of the disease and if $r>p$, we then remove in the future time
 $t_{j+1} + t_L + t_R$ this individual from the infectious condition and place
 him into the deceased condition $D(t)$.

 Along this paper, we will simulate the evolution of a disease possessing similar
 epidemiological parameters to those of the COVID-19. We use a basic reproduction
 number of $R_o = 4$, an incubation time of $t_L=4$ days and an infectivity time
 of $t_I=t_R-t_L=14$~\cite{Li20,Gu20,Ri20,Ta20,Sa20,Ba20,Ti20}.\\

 \section{\label{probinf} Modulation of the probability of infection}
 Without the inclusion of the weight function $W(t)$, the model we propose
 represents a probabilistic model with replacements, \ie, the probability of
 infecting a certain amount of susceptible per infected individual would be only
 determined by a stationary given value, independently of the total population
 being infected or the contact network, nevertheless, an intuitive
 behavior is that as the population of susceptible decreases, then also the
 chances of having large number of susceptible falling into close contact with
 the infectious population; in fact, this is exactly the underlying idea in the
 emergence of the herd immunity effect. On the other hand, the probability of
 infection is also continuously changing when contingency measures such as social
 distancing and home lockdowns are implemented in the population.
 In this regard, an appropriate functionality of the weight function could
 help us to incorporate effects such as herd immunity and contingency measures.

 To exemplify its effects to the epidemiological dynamics, we
 make use of a weight function which is the product of a function
 characterizing the herd immunity effect with an additional function
 characterizing the variability of the probability of infection along the
 transients of the epidemiological dynamics due to confinement and other
 related contingency measures, \ie:
 \begin{equation}
 W(t) =  H(t)\, C(t)
 \end{equation}
 where $H(t)$ represents the herd immunity effect which is estimated to emerge
 when a large proportion of the population (but not all), has gain certain
 immunity~\cite{TJ00,Rash12}, while $C(t)$ represents
 additional changes in the probability of infection along the transients of the
 dynamics. In the former, we employ a function in the form of a reversed
 logistic-like function whose argument depends on the fraction of the population
 that has become infected along the evolution of the epidemia, \ie:
 \begin{equation}\label{wf1}
 H(t)  = { 1+ \exp \left( {i(0)\over  \alpha } \right)
 \over 1+ \exp\left(   {  i(t)\over  \alpha}  \right) } (1-p) + p
 \end{equation}
 where $i(t_j) = \sum_{t =t_o}^{t_j} \sum_{i=1}^{I(t)} \chi_i(t) /N$, is the
 fraction of the cumulative infected population (Latent and Infectious) at time
 $t_j$; $\alpha$ is a free parameter serving to adjust the stationary value of
 the infection in the long time limit, and $p$ is a lower bound at which the
 probability of infection is reduced sufficiently to reach the stationarity.
 In our simulations, we set $p=0.1$, whereas we have seen that by choosing
 $\alpha=0.22$ the herd immunity is achieved when something close to the 80\% of
 the population has been infected~\cite{Rash12}.

\begin{figure}[!htb]
   \includegraphics[width=1\textwidth]{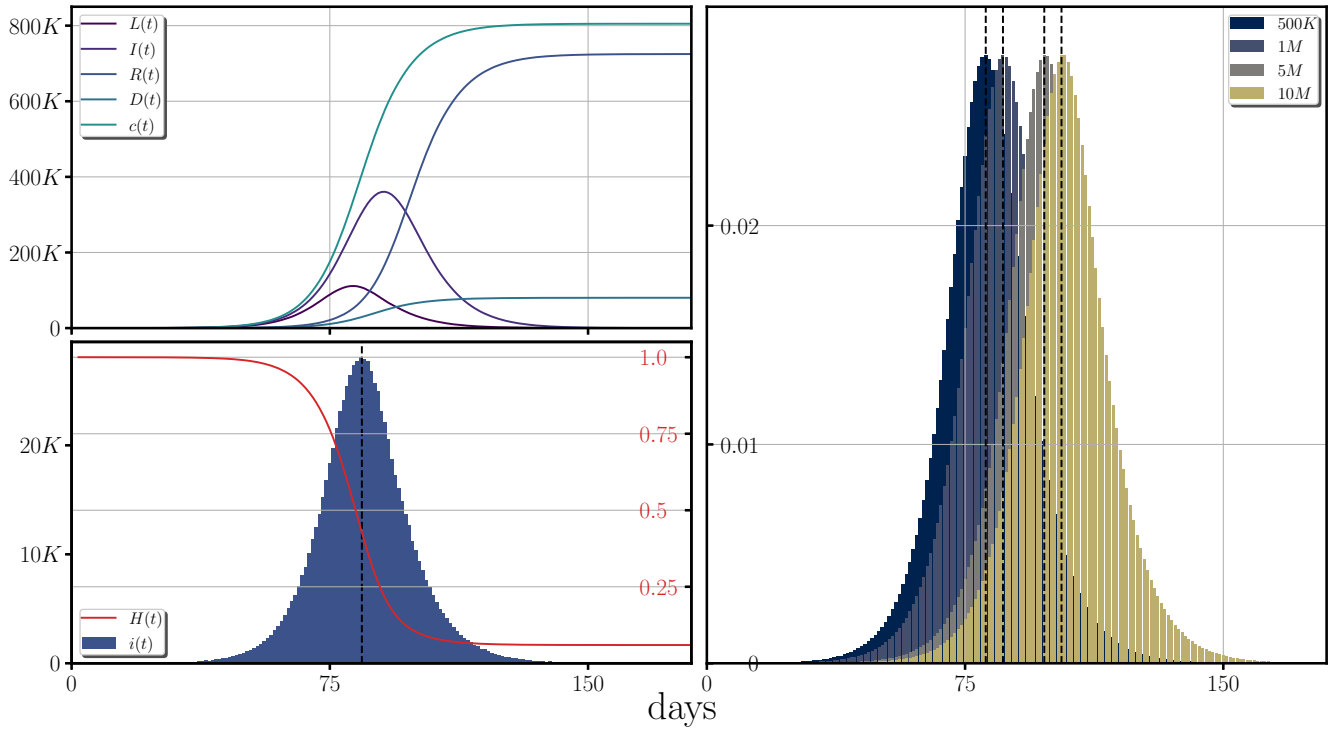}
 \caption{\label{fig1}
 Effects of the weight function representing the herd immunity
 (\ie, $W(t)=H(t)$), over the spreading of a disease based on the estimated
 COVID-19 parameters.
 At the first column (from left to right), the development of  the latent, the
 infectious, the recovered, the deceased population and the cumulative of the
 incidence are plotted for a single trajectory.
 In the next column is plotted the incidence with the form of the weight function
 representing the heard immunity effect while at the last column the averages
 over 5000 trajectories of the normalized incidence for different sizes of the
 population are shown. At the first row, from top to bottom, the simulations
 are generated when the fluctuations around the mean number of daily infections
 are generated through a punctual distribution \ie, no-fluctuations:
 $\Pi[\varrho_o] \rightarrow \{\lambda\}_i = \varrho_o \delta_{ii}$, while at the
 second row, the fluctuations are generated from a Gamma distribution with
 mean $k\theta = \varrho_o$ with a standard deviation
 of $\sigma = 1$, \ie $\Pi[\varrho_o]=\mathrm{Gamma}(k=\varrho_o,\theta=1)
 \rightarrow \{\lambda\}_i$. The epidemiological parameters employed in the
 figure corresponds to the estimated values of the COVID-19. }
 \end{figure}

 In figure \ref{fig1} we present, at the first column (from left to right),
 the effect of the weight function when it only characterizes the herd
 immunity effect (\ie~$W(t) = H(t),\; C(t)=1$), to the epidemiological variables,
 the incidence and its cumulative. In the  panels at the left
 we present the effect of the herd immunity of single realizations
 to the epidemiological variables (Latent, Infected, Recovered or deceased and
 the cumulative of the incidence ) and the incidence, together with the form
 of the weight function  generating the herd immunity effect. At the right panel,
 the effect of the herd immunity over the normalized incidence is presented for
 different  population sizes and when averaged over 5000 realizations.
 The parameters employed in figure \ref{fig1} are
 fixed to the estimated values of the COVID-19 disease described above.

 From figure \ref{fig1}, one notices that the maximum incidence for a population
 of one million is obtained of around 3 months after the beginning of the
 spreading of the disease, reaching at its maximum an amount of roughly
 2.5\% to 2.7\% of the total population. Additionally, if the population is
 increased in size by one order of magnitude, the maximum is shifted around 20 to
 30 days when no contingency measures are implemented in the population.
 The figure also tell us that for populations from one to ten millions the herd
 immunity may be reached between 130 to 160 days without confinement.


 On the other hand, the effect of non-pharmaceutical strategies implemented in
 the population to contain the spreading of a disease is another mean to modify
 the probability of infection. In this regard, one could think that some of
 the most common or intuitive
 responses of the population under an epidemiological risk: a confinement
 responding to the daily experience about the development of the disease, \eg,
 a confinement depending upon the number of active cases.
 In other words, when a certain fraction of the population has become
 symptomatic-infected (or deceased), it becomes more likely that some of the
 susceptible population has knowledge about infected individuals in their social
 circles or in the neighboring community, reacting with lockdowns due to the fear
 of becoming infected.
 Another possibility is that contingency measures are placed over the population
 (typically by health authorities) along different stages of the evolution of the
 disease, attempting (in principal) to find an equilibrium between the public
 health resources and different economical activities that requires contact among
 the population.
 In this case and as we have experienced with the COVID-19 pandemia,
 all populations have gone through lockdowns and
 relaxation of the confinements during different stages, which in turn,
 can be imposed at any time of the epidemiological development by the health
 authorities.

\begin{figure}[!htb]
   \includegraphics[width=1\textwidth]{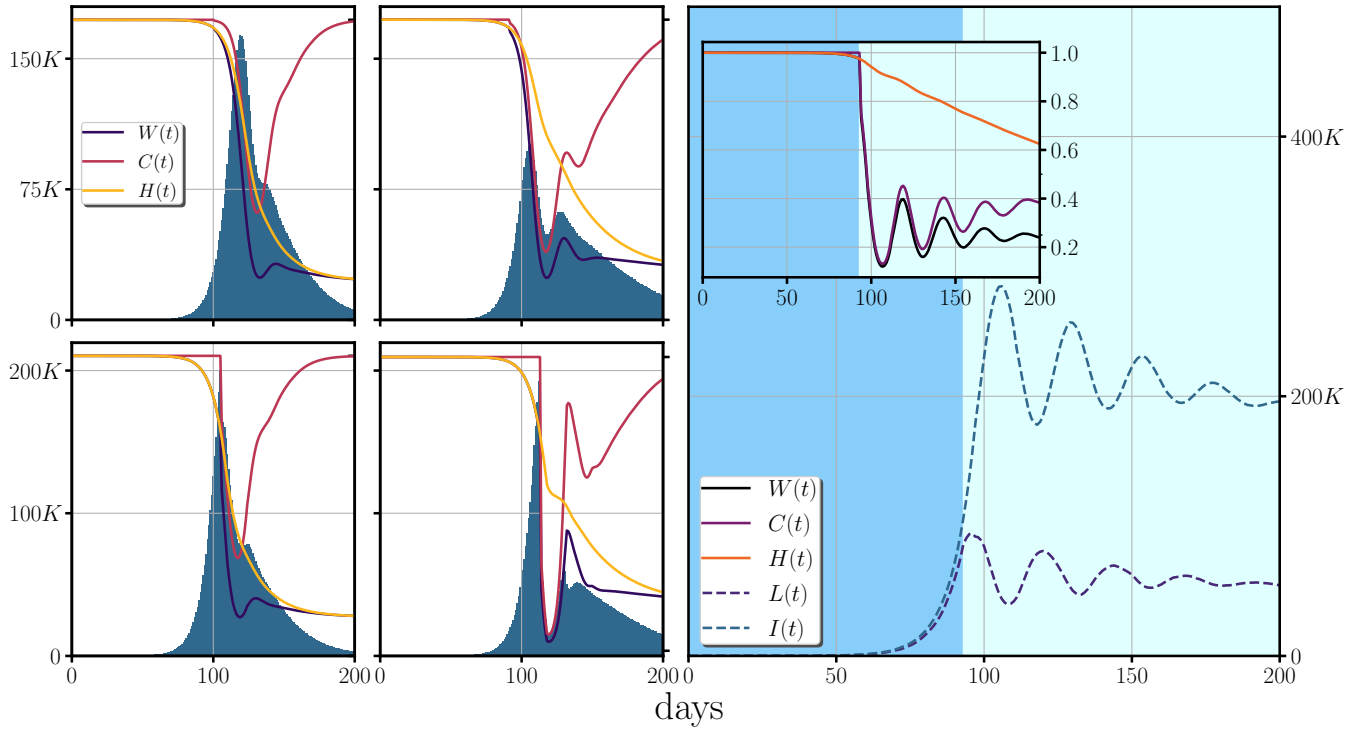}
 \caption{\label{fig2}
 The figure shows the effect of a confinement based on a Gaussian decay as given
 in (\ref{cgau}) for the COVID-19 parameters. At the four left panels, the
 incidence is shown with the weight function (containing the effects of herd
 immunity as presented before, and the effects of the confinement), over-imposed
 on the incidence. The rows from top to bottom show increasing values of $I_o$:
 $I_o=5\%$ (top), $I_o=10\%$ (bottom), while the columns from left to right show
 an increasing decaying rate parameter: $\gamma=5$ (left), $\gamma=10$ (right).
 At the right, the figure shows the effect of a strong and early confinement,
 $\gamma=50$ and $I_o=1\%$, to the latent and the infectious population. All
 cases corresponds to herd immunity parameters of $\alpha=0.22$ and $i_o=0$ and
 we use a total population of 10 million.}
 \end{figure}

 We use our stochastic model to explore the behavior of the COVID-19 spreading in
 two different confinement scenarios, a confinement triggered upon the number of
 infectious population (\ie the active cases) and confinement regulated by
 health authorities at different stages of the dispersion of the disease.
 In the former, we let
 $C(t)$ to be Gaussian-decaying function of the active cases, triggered once
 certain part of the population has become infectious-symptomatic or deceased,
 \ie:
 \begin{equation}\label{cgau}
   C(t) = 1 + \left[\exp(- [ \gamma \, I(t)/N ]^2 \,) - 1\right]\theta(I(t)-I_o)
 \end{equation}
 where $I(t)$ are the active cases at time $t$, while $I_o$ is a threshold
 telling the amount of active cases at which the confinement function is
 triggered; $\gamma$ is a decaying-rate parameter describing how strong is the
 confinement and $N$ is the total population.
 Figure \ref{fig2} shows the evolution of the disease for a confinement following
 a Gaussian decay as described in (\ref{cgau}). The left
 four frames represent $I_o=1\%$ and $I_o=10\%$ (from top to bottom) and
 $\gamma=5$ and $\gamma=10$ from the left to the right.

 In the figure one can sees that the outcomes of a confinement relying on the
 number of the infective population depends on how strong and rigorous is the
 confinement and at what stage of the dispersion of the disease is implemented.
 The different outcomes goes from a flattening of the epidemic curve, happening
 when the confinement does not happens abruptly, to revivals in the incidence
 which become periodic and more pronounced when the confinement is strong and
 happens at earlier stages of the epidemia.
 At the right panel we have plotted the latent and the infectious population when
 $I_o=1\%$ (\ie, the infectious individuals have reached one percent of the
 population) with an abrupt confinement, $\gamma=50$, from which several revivals
 can be seen. These revivals can be explained, by looking at the curve of the
 latent population: in an abrupt confinement, large part of the population
 remains on the latent condition and when the number of infectious population is
 reduced, the latent population will tend to break out the confinement measures
 beginning again with the contagious events. These results exhibit the need to
 employ correct times and duration of the confinement measures and that abruptly
 confinements without proper regulatory measures may trigger revivals.

\begin{figure}[!htb]
   \includegraphics[width=1\textwidth]{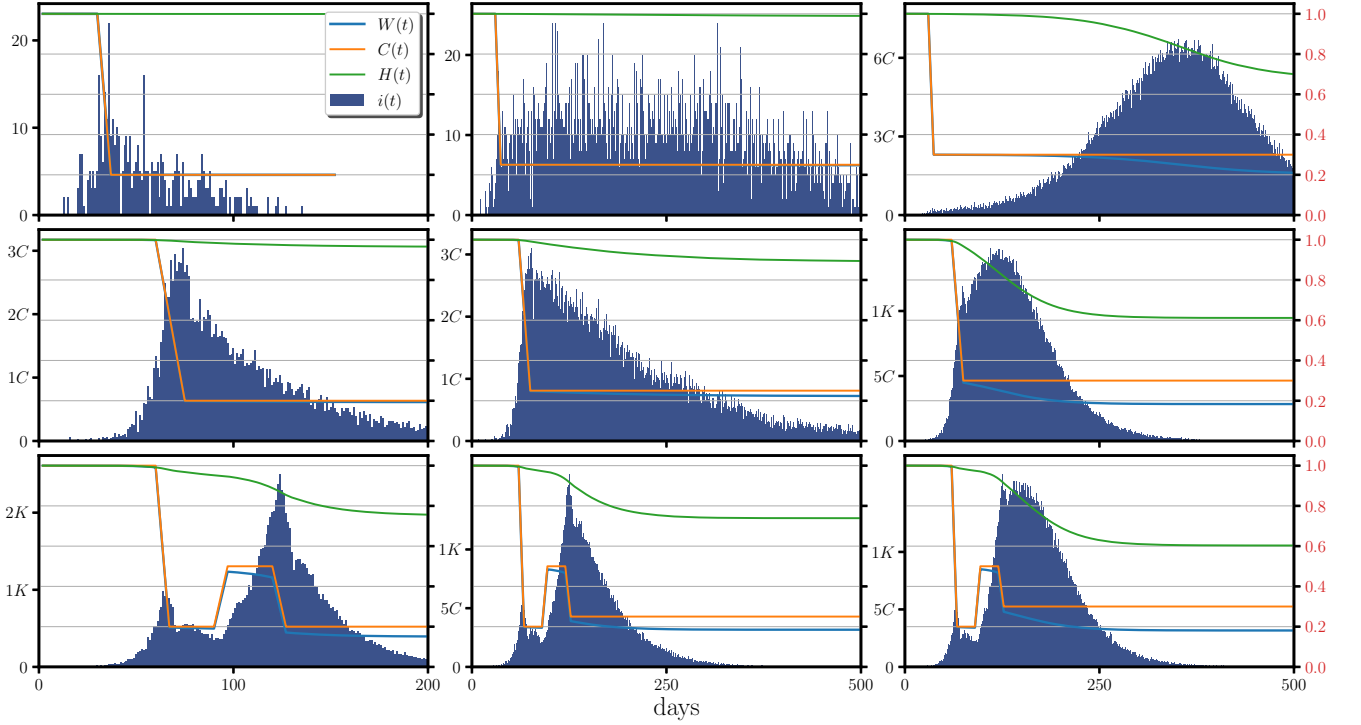}
 \caption{\label{fig3} The figure shows the effect of the piecewise confinement
 (plotted in the embedded small frames) over the incidence, using the
 COVID-19 parameters in a population of one million. At the first row (from top
 to bottom), the confinement begins at the day 30 and the decreasing along 7 days
 the function $C(t)$ from its initial value (one) to a 20\% of its initial value
 (first column), to a 25\% of its initial value (second column) and to a 30\%
 from its initial value (third column).
 At the second row the confinement begins at
 the day 60  where the function $C(t)$ is decreased from its initial value
 to a 20\% of its initial value (first column), to a 25\% of its
 initial value (second column) and to a 30\% from its initial value (third
 column). Finally at the third row the function $C(t)$  is initially decreased to
 a 20\% of its initial value along a period of 7 days, increased back
 to a value of a 50\% of its initial value along a period of
 7 days and finally decreased back again at the day 120 along
 a period of 7 days to a 20\% of its initial value (first column), to a 25\%
 (second column) and to a 30\% (third column).}
 \end{figure}

 In the context of a confinement based on regulatory measures such as lockdowns,
 social distancing and restrictions on mobility; they could be implemented
 at any time of the epidemiological development, it is evident that they will not
 follow a deterministic behavior (as shown previously).
 In this regard, in figure \ref{fig3}, we explore the behavior of the
 spreading of the disease when the probability of infection is manipulated by a
 piecewise time-dependent $C(t)$ function.
 In this figure, we show single realizations of the behavior of
 the incidence and one can sees that if confinement is applied at
 relatively early stages, then a reduction of the $C(t)$ function below the
 25\% of its initial value produces a deceleration of the incidence, at 25\% the
 incidence is maintained approximately at a constant rate while anything above
 the 25\% will corresponds to increments in the incidence with stronger
 accelerations for larger values of the $C(t)$ function.
 In this regard, figure \ref{fig3}
 tell us that it is possible to accelerate, decelerate or keep steady the
 incidence of a disease with an appropriate manipulation of the weight
 function.
 In section \ref{fcast} we use this acquired knowledge
 to fit the beginning of the spreading of the COVID-19 to real cases happening in
 Mexico.

 \section{\label{wfint}Interpretation of the weight function}
 As shown before, the employment of the weight function serves to modulate
 the mean of the daily infected population and hence it must posses a relation
 in describing variations of the probability of infection in populations
 following structured behavior. In this regard, one may ask about the
 relation between $W(t)$ and the effective reproduction number $\mathcal{R}(t)$,
 the latter representing the statistical mean of the infections caused by single
 individuals once the disease has begun to disperse in the population.
 To answer this question, lets consider first the statistical mean of the
 number of infected at the time $t_j$ due to the $i$-th infectious individual:
 \begin{equation}
   N_i(t_j) = \sum_k[\chi_i]_k\, p_k[\lambda_i W(t_j)] = \lambda_i W(t_j)
 \end{equation}
 where the $[\chi_i]_k$ represents the possible outcomes of the random variable
 $\chi_i$ of the $i$-th infectious at time $t_j$, \ie, the possible number of
 infections that the $i-th$ infectious could produce with probability
 $p_k[\lambda_i W(t_j)]$ of the $k$-th event and $\lambda_i W(t_j)$ is the
 statistical mean of all possible outcomes of the $i$-th contagious individual
 at time $t_j$. In the case where the Poisson parameters $\lambda_i$ are also
 distributed according to a probability distribution $\Pi[\varrho_o]$
 (see section \ref{model}), then the average of the total number of infected
 individuals at a fixed time $t_j$ may therefore be given by:
 \begin{equation}
   \bar{N}(t_j) =  \sum_{i=1}^{I(t_j)} { N_i(t_j)\over I(t_j) } = W(t_j) \,
   \sum_{i=1}^{I(t_j)} {\lambda_i\over I(t_j)}.
 \end{equation}

 Let us consider now that in certain given time $t_j\geq t'$, the number of
 infectious has become large and representative about the dispersion of the
 disease in the population, \ie, the number of infectious can be found
 homogeneously distributed in the population. This should happen after certain
 time when all the possible outcomes of the set of the random variables
 $\lambda_i$ are centered around the mean of the distribution $\Pi[\varrho_o]$.
 In other words, for times $t<t'$ fluctuations are expected to dominate and as
 the number of infectious increases, the fluctuations reduce yielding a more
 localized value of the probability of infection. Therefore, at time $t_j> t'$
 one could approach the number of infected due to the infectious as
 $\sum_{i=1}^{I(t_j)} \lambda_i / I(t_j) \sim \varrho_o$, yielding us a close
 relation between the weight function defined earlier and the time-dependent
 effective reproduction number:
 \begin{equation}\label{Rt}
 \mathcal{R}(t_j) = \bar{N}(t_j) t_I  \sim R_o \, W(t_j)\,.
 \end{equation}

 \subsection{Empirical estimation of $\mathcal{R}(t)$.}
 As seen before, the weight function could be interpreted as a normalization of
 the  time-dependent effective reproduction number, \ie
 $W(t) = \mathcal{R}(t)/R_o$.
 In this context, having a self-consistent mechanism that could provide us with
 an estimation about the evolution of the effective reproduction number based on
 the real available data, would be desirable.
 We approximate to this problem by using the information accessible through
 empirical data, such as the empirical incidence $i_e(t)$ and its cumulative
 $c_e(t)$. In our stochastic approximation, the daily synthetic incidence is
 obtained from a set of random variables  following a Poisson distribution;
 \ie $i_s(t_j) = \sum_{j=1}^{I(t_j)}\chi_i(t_j)$, hence the statistical mean of
 the cumulative of the daily incidence may be written as:
 \begin{equation}
   \bar{c}_s(t_j) = {1\over M}\sum_{l=1}^M \sum_{k=1}^{j}
   \sum_{i=1}^{I_l(t_k)}\chi^{(l)}_i(t_k)
 \end{equation}
 where $M$ represents the total number of trajectories to which the statistical
 mean is performed.  If the number of trajectories is large enough, then the
 statistical mean will approach the expectation value of the random variables,
 \ie, the Poisson parameters associated to  the infectious individuals.

 By considering the average over the ensemble of the infectious individuals,
 we can approximate the average of the cumulative as:
 \begin{eqnarray}\nonumber
 \bar{c}_s(t_j) &\approx&
 \sum_{k=1}^{j}\sum_{i=1}^{\bar{I}(t_k)} \lambda_i W(t_k)
 = \sum_{k=1}^{j}\sum_{i=1}^{\bar{I}(t_k)} \lambda_i \mathcal{R}(t_k)/R_o\\ \,,
 \end{eqnarray}
 where in the last line, we have done the replacement of the weight function
 through the relation obtained earlier in (\ref{Rt}). Furthermore, in order to
 gain convergence, we fix again the Poisson parameters $\lambda_i$ to be obtained
 from a punctual distribution, \ie
 $\Pi[\varrho_o]\rightarrow \{\lambda\}_i = \varrho_o \delta_{ii} =
 R_o/t_I \,\delta_{ii}$, hence we write for the average of the cumulative:
 \begin{equation}\label{avcum}
   \bar{c}_s(t_j) \approx
   \sum_{k=1}^{j} \bar{I}(t_k) \mathcal{R}(t_k)/t_I\,.
 \end{equation}
 Our aim is to give an approximate description about the time dependent
 reproduction number through empirical quantities; in this regard, we do the
 replacement of the average synthetic cumulative and the averaged infectious
 population with their correspondent empirical descriptions;
 $\bar{c}_s(t_j)\rightarrow c_e(t_j)$ and $\bar{I}(t_j)\rightarrow I_e(t_j)$.
 Additionally by expanding the sum in (\ref{avcum}) to the first step of
 propagation and obtaining recurrently the average of the cumulative, is easy to
 see that the time dependent reproduction number can be obtained as:
 \begin{equation}
   \mathcal{R}(t_j) = {i_e(t_j)\over I_e(t_j)} t_I\,,
 \end{equation}
 while the number of infectious individuals at time $t_j$ can be approximated,
 according the definitions done earlier, as:
 $I_e(t_j) = \sum_{k=0}^{t_I} i_e( j - t_L - k)$.

 \section{\label{fcast} Development of the COVID-19 in some Mexican states.}

 \begin{figure}[!htb]
   \includegraphics[width=1\textwidth]{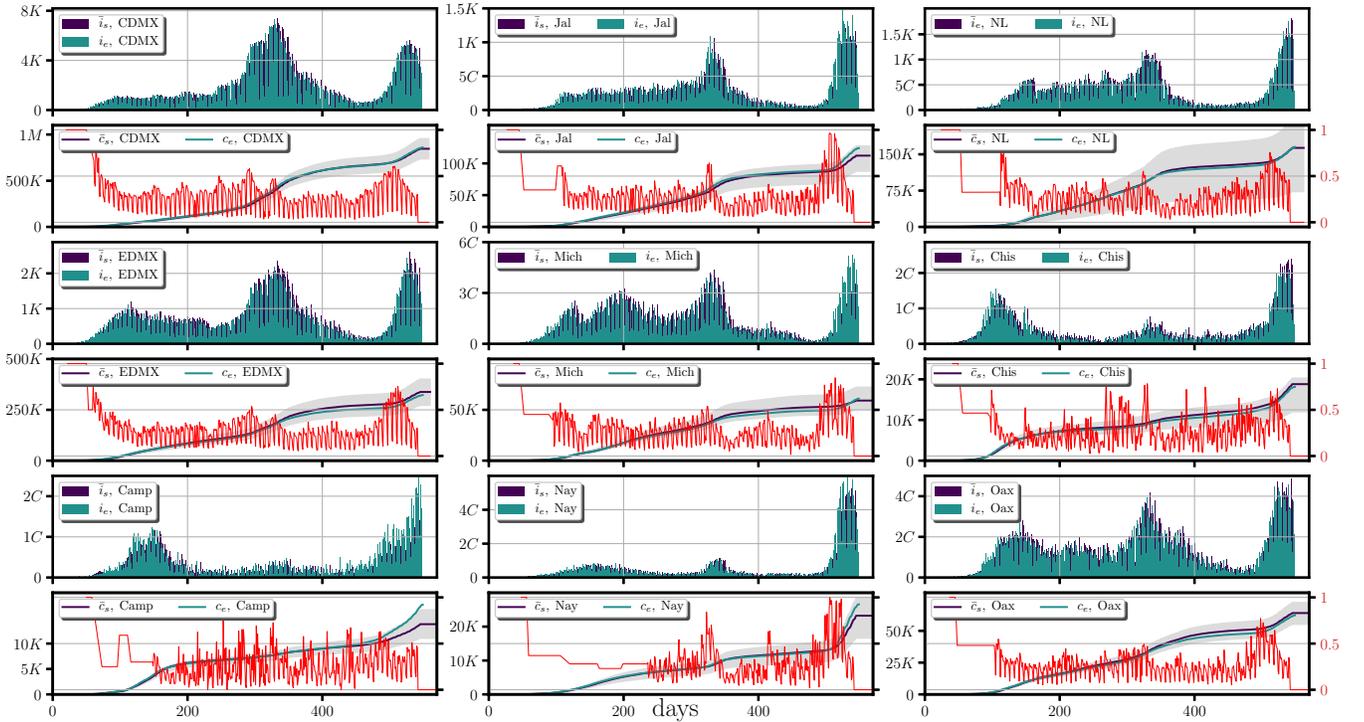}
 \caption{\label{fig4} The figure shows a comparison of the incidence
 and its cumulative, between the synthetic data
 generated by the stochastic model when averaged over 5000 trajectories to the real
 scenarios haopening in some Mexican states and Mexico City. The synthetic
 data was generated by employing the empirical estimation of the effective
 reproduction number from which the weight function can be obtained:
 $W(t) = \mathcal{R}(t)/ R_o $. The figure shows a period of roughly a year and a
 half of the spreading of the SARS CoV-2 (from February the 18th of 2020 to
 August the 20th of 2021).}
 \end{figure}

 Along the development of the COVID-19 pandemia, we have used the stochastic
 model to follow the evolution of the spreading of the COVID-19 in certain
 Mexican states, some of them housing some of the largest Mexican metropolitan
 areas (\ie, México City, Estado de México, Jalisco and Nuevo León) and some
 middle-size populated states.
 To model the spreading of the COVID-19 in these states,
 we begin by looking to the empirical incidence and consider an initial infectious
 estimation by adding the number of the daily new infected population until
 a the day in which the following days a continuous incidence is sustained
 (\ie~there is at least one new infected person reported); afterwards,
 we perform rough estimations about the form of the weight function following
 linear decrements (deceleration) or linear increments (acceleration) at certain
 intervals of time to describe the development of the COVID-19 at the beginning
 of the spreading, and once the number of infectious individuals has become
 sufficiently large (in this case we wait until the number of infectious
 individuals has grown beyond 1000 individuals for the first time), we employ the
 empirical estimation about the effective reproduction number to describe the
 evolution of the pandemia in these Mexican states. In these simulations, we only
 employ the empirical estimations about the effective reproduction number without
 considering herd immunity effects since under confinement, the pandemia can be
 considered effectively open in the sense that there is an infinite number of
 susceptible.
 The real data has been obtained from the reported cases by the scientific
 division of the  Mexican federal government (CONACyT):
 \texttt{https://datos.covid-19.conacyt.mx/}.
 These results are presented in figure \ref{fig4} for the cumulative of the
 incidence when averaged over 5000 trajectories.
 In figure \ref{fig4} one can notice, in the general,
 a good agreement between the synthetic data generated from the stochastic model
 and the real scenarios along the period of dispersion of the
 disease which was considered of roughly a year and a half. Furthermore,
 the good agreement indicates that the infections events can be assumed
 to be time-homogeneously distributed, except for some cases at
 certain times (\eg~see Campeche around the 450 day) which may suggest the
 occurrence of anomalous events (possibly superspreading events) for which this
 assumption may not hold.

 \begin{figure}[!htb]
   \includegraphics[width=1\textwidth]{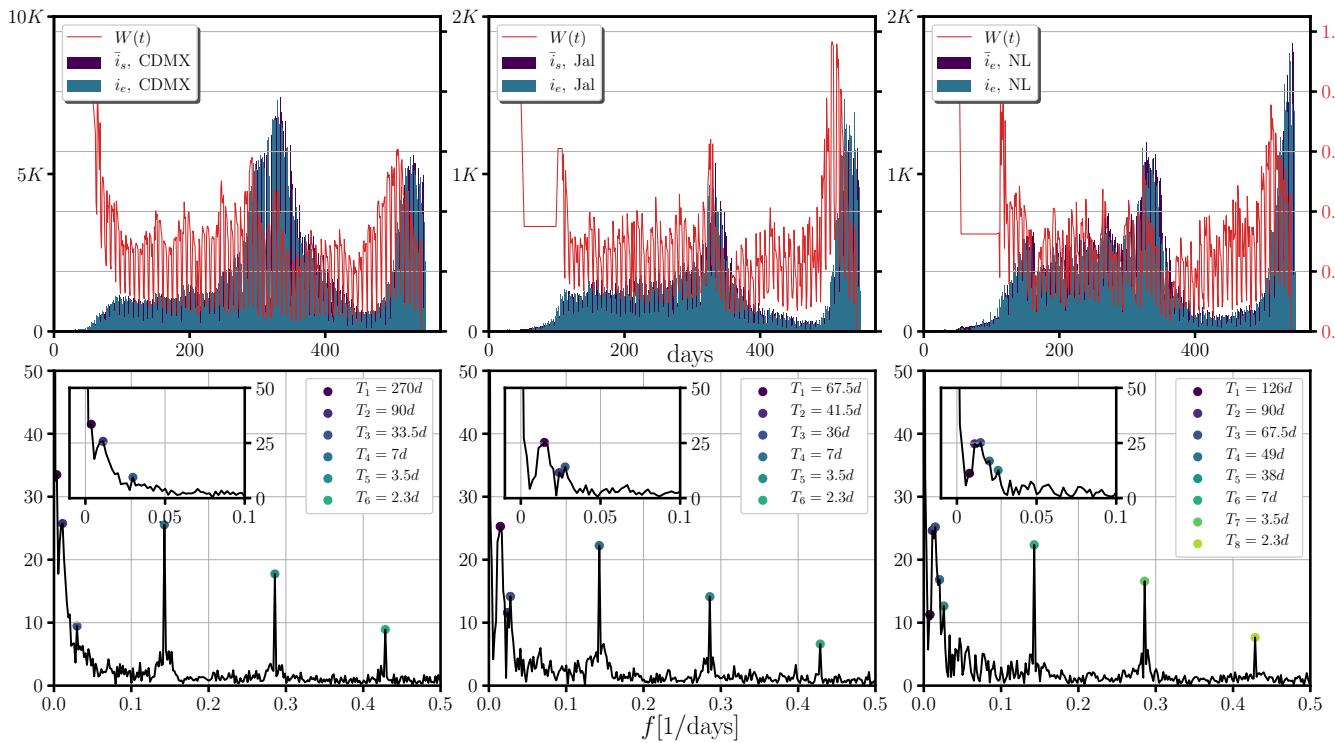}
  \caption{\label{fig5a}
 The figure shows the absolute value of the one-sided Fourier transform of the
 probability of infection derived from the empirical effective reproduction
 number of Mexico City and the states of Jalisco and Nuevo León. At the first row
 the empirical and synthetic incidence
 and the form of the weight function are shown as reference. At the
 second row the absolute value of the one-sided Fourier transform of the weight
 function is shown, marking the most relevant frequencies associated to
 periodic patterns in the effective reproduction number.
 The figure shows a
 period of roughly a year and a half of the spreading of the SARS-CoV-2 (from
 February the 18th of 2020 to August the 20th of 2021).}
 \end{figure}

 \begin{figure}[!htb]
   \includegraphics[width=1\textwidth]{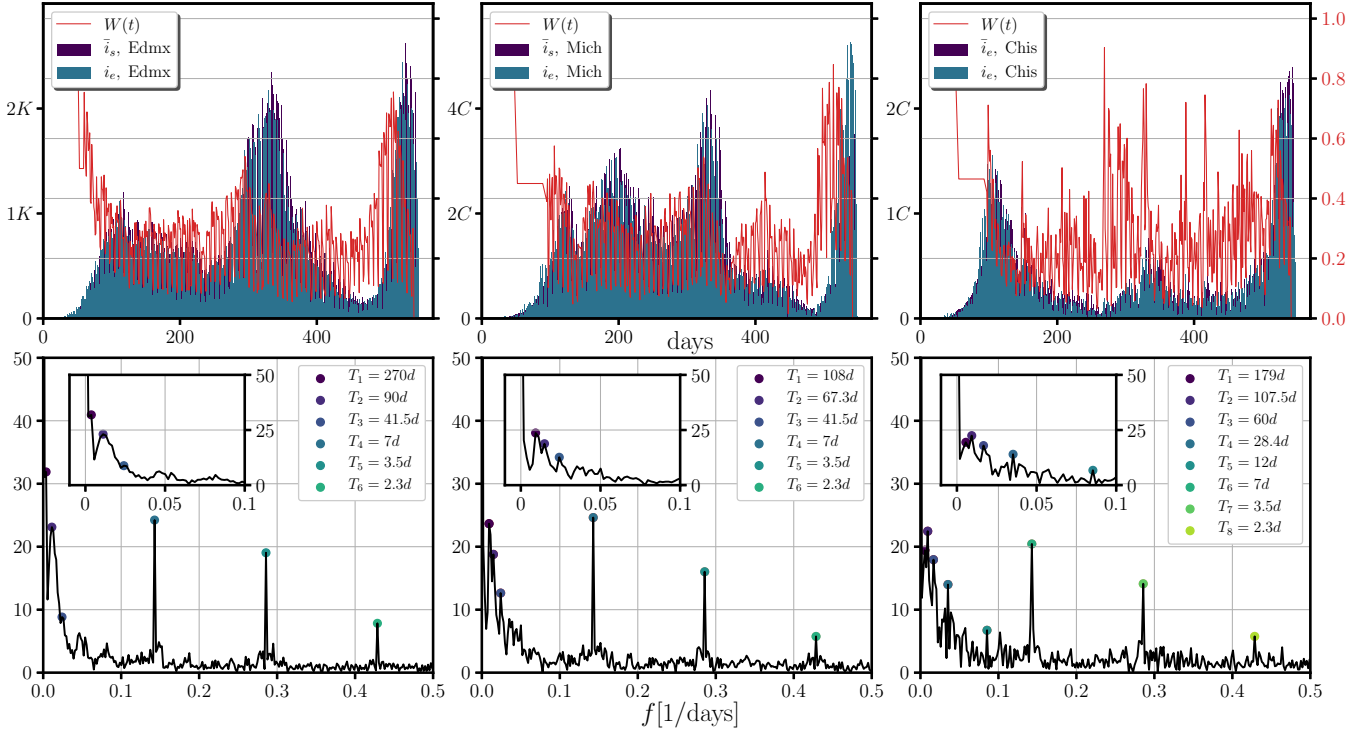}
  \caption{\label{fig5b}
 The figure shows the absolute value of the one-sided Fourier transform of the
 probability of infection derived from the empirical effective reproduction
 number of the states of Estado de México of Michoacan and Chiapas.
 At the first row the empirical and synthetic incidence and the form of the
 weight function are shown as
 reference. At the second row the absolute value of the one-sided Fourier
 transform of the weight function is shown, marking the most relevant frequencies
 associated to periodic patterns in the effective reproduction number.
 The figure shows a
 period of roughly a year and a half of the spreading of the SARS-CoV-2 (from
 February the 18th of 2020 to August the 20th of 2021).}
 \end{figure}

 \begin{figure}[!htb]
   \includegraphics[width=1\textwidth]{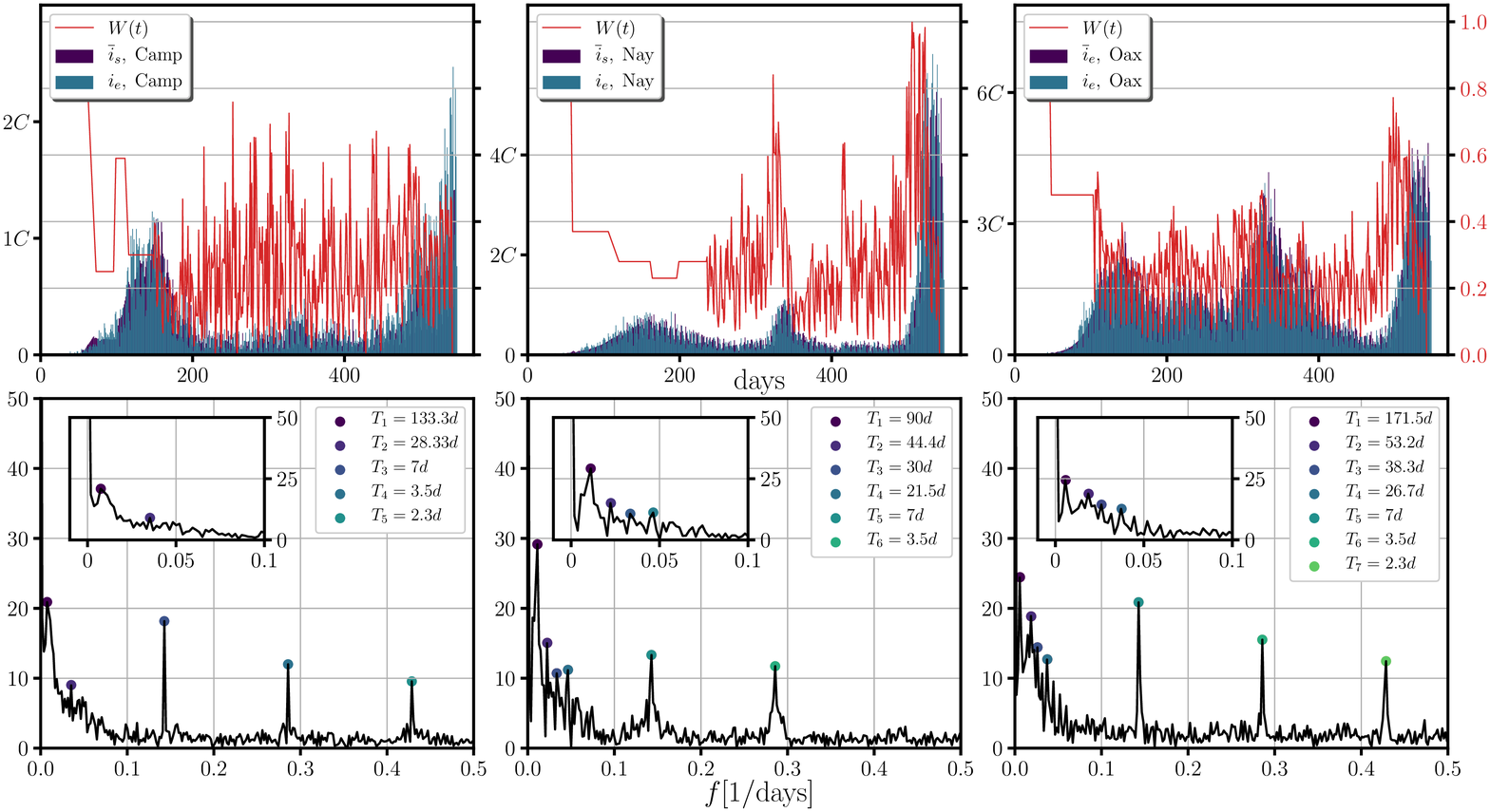}
  \caption{\label{fig5c}
 The figure shows the absolute value of the one-sided Fourier transform of the
 probability of infection derived from the empirical effective reproduction
 number of the states of Campeche of Nayarit and Oaxaca.
 At the first row the empirical and synthetic incidence and the form of the
 weight function are shown as
 reference. At the second row the absolute value of the one-sided Fourier
 transform of the weight function is shown, marking the most relevant frequencies
 associated to periodic patterns in the effective reproduction number.
 The figure shows a
 period of roughly a year and a half of the spreading of the SARS-CoV-2 (from
 February the 18th of 2020 to August the 20th of 2021).}
 \end{figure}

 Once we have confirmation about the good agreement of the model,
 we turn our attention to the empirical weight function
 (\ie~the normalized effective reproduction number) and look for periodic patterns
 in the daily mean of infections. The
 period of these patterns can be obtained by performing the Fourier transform of
 the weight function where the peaks appearing in the spectrum of the
 weight function will corresponds to the frequencies associated to
 these periods and could provide us with information about
 the implementation of confinement, testing, mobility, or
 even emergent periodic behavior encoded in the probability of infection,
 during the development of the COVID-19.

 In figures \ref{fig5a} to \ref{fig5c} we present the absolute value of the
 one-sided Fourier transform of the weight function (second row) of each of the
 Mexican states analyzed before (first row).
 In the figure we identify three set of frequencies whose corresponding
 periods lies on different time scales which we believe are related to the
 weekly agenda, confinements and
 also to larger patterns such as pandemic waves or even the
 emergence of seasonality. We identify a first
 set belonging to periods within a week
 for which in almost all the cases (with the exception of Nayarit, which in
 addition, it has been one of the states with the smaller incidence rate before
 the third pandemic wave) three peaks are present exactly at the same periods
 \ie~at $7$ days, $3.5$ days and $2.3$ days which may confirm
 a global (not local) behavior and
 we believe these are related to the weekly agenda concerning the periodicity of
 the testing and the capture of the readouts. A second
 time scale lies on the range from one two six months which we believe
 are connected to pandemic waves or periods of confinement, de-confinement and
 holiday seasons. For this time scale one observes periods that are shared by
 certain states (see \eg~ the periods of 90 days appearing in México City,
 Estado the México, Nuevo León and Nayarit, or the state of Jalisco and
 Michoacan which share periods of 67.5 days and 41.4 days, the latter also
 shared by Estado de México) and additional periods which are unique of
 the particular cases.
 The shared periods could be explained due to the closeness of certain states
 (such is the case of México City and Estado the México or Jalisco and Michoacan)
 or due to the connections of large populated regions such as Nuevo León with
 México city and Estado de México.
 The larger time scale is
 represented by the first peak appearing in México City and Estado de México which
 is not surprising they shared this peak since they large neighboring metropolitan
 areas. This peak corresponds to a periodic pattern emerging at $T_1=297$ days,
 suggesting which may be related to pandemic waves or even the emergence of a
 seasonal behavior of the COVID-19.
 This time scale is only present in these states and it may be consequence of
 the amount of the population that has become infected (notice
 that in Mexico City the amount of infected population is roughly one order of
 magnitude greater than the states of Jalisco and Nuevo León), and the large
 degree of urbanization of this region.

 \section{\label{con} Conclusion}
 In this paper we have derived an stochastic compartmental epidemiological model
 constructed from first principles which consist on a randomization about the
 number of the new infected population caused daily and by assuming that the
 infectious events follow a Poisson process.
 We have shown that under this assumption, one can reconstruct and simulate
 emergent phenomena such as herd immunity or certain confinement scenarios, by
 introducing an additional time-dependent function (the weight function) which
 can be represented as a normalization of the time-dependent effective
 reproduction number which additionally has helped us to modulate the mean of
 the daily new infections, (and hence the
 probability of infection) along the evolution of the disease.
 Moreover, we have derived an empirical estimation about the weight
 function which has serve us to incorporate self-consistency into the model.
 Along this paper, we have focused
 on the epidemiological parameters corresponding to the COVID 19 disease
 and through the employment of the weight function, the model is capable
 to introduce and study some conceptual behaviors such as
 herd immunity or certain idealized confinement scenarios. In the former
 we have employed an inverse-like logistic
 function of the fraction of the total infected population, which for the
 COVID-19 epidemiological parameters and without any confinement measures, we
 have found that the peak of the incidence scales by 20 to 30 days when the total
 population is increased by one order of magnitude while independently of the
 population sizes, the maximum incidence reaches a reaching 2.5\% to 2.7\% of the
 total population; in the latter, we have explored the
 reaction of the dispersion of the disease when the population reacts to the
 infectious population (representing an intuitive reaction of the population
 under an epidemiological emergence), finding revivals in the incidence
 (infective waves) if confinement is abrupt and happens at earlier stages in the
 dispersion of the disease, and a flattening of the epidemic curve on the
 contrary situation.
 Furthermore, we have simulated the effects on the incidence when the weight
 function is described through a piecewise function finding accelerations in the
 incidence when the weight function takes values above 25\% of its  initial
 value, a steady behavior for a 25\% of its initial value and deceleration in
 the incidence when the weight function take values below the 25\% of its
 initial value.

 We have employed our stochastic model model together with
 the definition of the empirical  weight function, to simulate
 the dispersion of the COVID-19 in some Mexican states, some of them
 housing the major metropolitan areas in Mexico and finding a very good agreement
 to the real scenarios implying that the infectious events in México could be
 interpreted as homogeneously distributed events.

  Finally, we have applied the one-sided Fourier transform to the empirical
  description of the weight function with the intention to look at periodic
  patterns emerging in the mean of the daily infection which may give us
  insights about the evolution of the pandemia in México. In this regard,
  we have found three different set of frequencies corresponding to
  different time-scales which we identify to a weekly agenda about
  the capture of the readouts of the testings, confinement and also larger
  patterns which may be related to pandemic waves or even seasonality.

\acknowledgments
PCL acknowledges financial funding from CONACyT through the research
project Ciencia de Frontera 2019 (No. 10872). Also all the
authors acknowledges the very helpful discussion with Thomas Gorin,
Soham Biswas, Ulises Moya and Raúl Nanclares.

\appendix
\section*{\label{app1} The standard SIR model limit}
In the context of the standard SIR model, the susceptible population $S$
has to be considered corresponding to the part of the population
which can be infected. In this regard, its number is constantly reducing due to
the contagious events, therefore, according to the random
scenario we have proposed, at the time $t_{j+1}$ the number of susceptible
can be described as:
\begin{equation}\label{ns}
  S(t_{j+1}) =  S(t_{j}) - \sum_{i=1}^{I(t_{j})} \chi_i(t_{j})\,.
\end{equation}
with the last term accounting for the part of the susceptible population
skiping out from this conditon into the latent condition.

From our random model, the standard SIR model can be derived by
considering that at any fixed time $t_j$, any infectious individual
infects the same amount of susceptible. This assumptions can be considered
valid, (althouhg unrealistic somehow), when the population is homogenously
distributed along the infection area
at all time, \ie, there are no clusters of individuals in the population and the
reorganization of the susceptible popultion every time step is homogeneously
distributed in space. In this regard, the SIR model limit relies on the
assumptions that the distribution of the individuals of the population is
independent of the antropological characteristics of the society.

Following within this idea, we do the replacement of the new daily infected
population by each infectious individual by a constant ammount $S_o(t_j)$; the
total amount of new infected population at the time $t_j$ will yield:
\begin{eqnarray}\nonumber
  \sum_{i=1}^{I(t_j)} \chi_i(t_{j}) &= &\sum_{i=1}^{I(t_j)} S_o(t_j)\\
   & = & S_o(t_j)\, I(t_j)\,.
\end{eqnarray}

Moreover, under this assumption, the number of the daily new infected
population is scalable in time, hence the population of susceptible can be
thought as reservoir of individuals of infinite size such that the epidemic
events will not alter the homogenity of the distribution and
one can connect the number of susceptible per unit time
$\delta_t = t_j - t_{j-1}$, to the total number of susceptible in the long time
limit, \ie ~ $ S_o (t_j)/\delta_t = S(t_j)/ n \delta_t $.
Finally, by redefining the number of the infected population as
$L+I \rightarrow I$ and by considering the limit of infinitesimal time steps
(which will be  valid only for very large population sizes) then, one can write
now the set of deterministic equations of the SIR model about the evolution of
the disease as:

\begin{eqnarray}\label{nsas}
\dot{S}(t) &=& -  \beta\, S(t)\,I(t)  ,\\\label{nias}
\dot{I}(t)  &=&  \beta \, S(t)\, I(t) - \kappa_R\,I(t)\,,\\\label{nras}
\dot{R}(t) &=& \kappa_R\,I(t)\,.
\end{eqnarray}
where contact rate defined as the average number of contacts per
individuals per time will be given by $\beta  =
\lim_{\delta_t\rightarrow 0}
( n  \delta t )^{-1}$ while the recovery rate will fulfill: $\kappa_R =
\lim_{\delta_t\rightarrow 0 }(r \delta_t)^{-1}$,
with $r$ being a recovery-time scale factor, \ie $t_R = r \delta_t$.
Clearly, for both rates as $\delta_t \rightarrow 0$ the quantities
$n\delta t$ and $r \delta t$ remain finite.

\section*{Additional information}
\textbf{Accession codes}:\texttt{https://github.com/RenatoSalArrDu/StochasticSLIRD}\\

\bibliography{man}

\begin{thebibliography}{30}%
\makeatletter
\providecommand \@ifxundefined [1]{%
 \@ifx{#1\undefined}
}%
\providecommand \@ifnum [1]{%
 \ifnum #1\expandafter \@firstoftwo
 \else \expandafter \@secondoftwo
 \fi
}%
\providecommand \@ifx [1]{%
 \ifx #1\expandafter \@firstoftwo
 \else \expandafter \@secondoftwo
 \fi
}%
\providecommand \natexlab [1]{#1}%
\providecommand \enquote  [1]{``#1''}%
\providecommand \bibnamefont  [1]{#1}%
\providecommand \bibfnamefont [1]{#1}%
\providecommand \citenamefont [1]{#1}%
\providecommand \href@noop [0]{\@secondoftwo}%
\providecommand \href [0]{\begingroup \@sanitize@url \@href}%
\providecommand \@href[1]{\@@startlink{#1}\@@href}%
\providecommand \@@href[1]{\endgroup#1\@@endlink}%
\providecommand \@sanitize@url [0]{\catcode `\\12\catcode `\$12\catcode
  `\&12\catcode `\#12\catcode `\^12\catcode `\_12\catcode `\%12\relax}%
\providecommand \@@startlink[1]{}%
\providecommand \@@endlink[0]{}%
\providecommand \url  [0]{\begingroup\@sanitize@url \@url }%
\providecommand \@url [1]{\endgroup\@href {#1}{\urlprefix }}%
\providecommand \urlprefix  [0]{URL }%
\providecommand \Eprint [0]{\href }%
\providecommand \doibase [0]{http://dx.doi.org/}%
\providecommand \selectlanguage [0]{\@gobble}%
\providecommand \bibinfo  [0]{\@secondoftwo}%
\providecommand \bibfield  [0]{\@secondoftwo}%
\providecommand \translation [1]{[#1]}%
\providecommand \BibitemOpen [0]{}%
\providecommand \bibitemStop [0]{}%
\providecommand \bibitemNoStop [0]{.\EOS\space}%
\providecommand \EOS [0]{\spacefactor3000\relax}%
\providecommand \BibitemShut  [1]{\csname bibitem#1\endcsname}%
\let\auto@bib@innerbib\@empty
\bibitem [{\citenamefont {{World Health
  Organisation}}(2021{\natexlab{a}})}]{WHO1}%
  \BibitemOpen
  \bibfield  {author} {\bibinfo {author} {\bibnamefont {{World Health
  Organisation}}},\ }\href@noop {} {} (\bibinfo {year} {2021}{\natexlab{a}}),\
  \bibinfo {note} {\url{https://covid19.who.int/}, Last accessed on
  2021-09-09}\BibitemShut {NoStop}%
\bibitem [{\citenamefont {Shankar}\ \emph {et~al.}(2021)\citenamefont
  {Shankar}, \citenamefont {Mohakuda}, \citenamefont {Kumar}, \citenamefont
  {Nazneen}, \citenamefont {Yadav}, \citenamefont {Chatterjee},\ and\
  \citenamefont {Chatterjee}}]{Shan21}%
  \BibitemOpen
  \bibfield  {author} {\bibinfo {author} {\bibfnamefont {S.}~\bibnamefont
  {Shankar}}, \bibinfo {author} {\bibfnamefont {S.~S.}\ \bibnamefont
  {Mohakuda}}, \bibinfo {author} {\bibfnamefont {A.}~\bibnamefont {Kumar}},
  \bibinfo {author} {\bibfnamefont {P.}~\bibnamefont {Nazneen}}, \bibinfo
  {author} {\bibfnamefont {A.~K.}\ \bibnamefont {Yadav}}, \bibinfo {author}
  {\bibfnamefont {K.}~\bibnamefont {Chatterjee}}, \ and\ \bibinfo {author}
  {\bibfnamefont {K.}~\bibnamefont {Chatterjee}},\ }\href {\doibase
  10.1016/j.mjafi.2021.05.005} {\bibfield  {journal} {\bibinfo  {journal}
  {Medical journal, Armed Forces India}\ }\textbf {\bibinfo {volume} {77}},\
  \bibinfo {pages} {S385—S392} (\bibinfo {year} {2021})}\BibitemShut
  {NoStop}%
\bibitem [{\citenamefont {Weiss}(2013)}]{Hy13}%
  \BibitemOpen
  \bibfield  {author} {\bibinfo {author} {\bibfnamefont {H.}~\bibnamefont
  {Weiss}},\ }\href {http://mat.uab.cat/web/matmat/es/v2013n03/} {\bibfield
  {journal} {\bibinfo  {journal} {Materials Matem\`atics}\ }\textbf {\bibinfo
  {volume} {3}},\ \bibinfo {pages} {17} (\bibinfo {year} {2013})}\BibitemShut
  {NoStop}%
\bibitem [{\citenamefont {Read}\ \emph {et~al.}(2021)\citenamefont {Read},
  \citenamefont {Bridgen}, \citenamefont {Cummings}, \citenamefont {Ho},\ and\
  \citenamefont {Jewell}}]{Re21}%
  \BibitemOpen
  \bibfield  {author} {\bibinfo {author} {\bibfnamefont {J.~M.}\ \bibnamefont
  {Read}}, \bibinfo {author} {\bibfnamefont {J.~R.~E.}\ \bibnamefont
  {Bridgen}}, \bibinfo {author} {\bibfnamefont {D.~A.~T.}\ \bibnamefont
  {Cummings}}, \bibinfo {author} {\bibfnamefont {A.}~\bibnamefont {Ho}}, \ and\
  \bibinfo {author} {\bibfnamefont {C.~P.}\ \bibnamefont {Jewell}},\ }\href
  {\doibase 10.1098/rstb.2020.0265} {\bibfield  {journal} {\bibinfo  {journal}
  {Philosophical Transactions of the Royal Society B: Biological Sciences}\
  }\textbf {\bibinfo {volume} {376}},\ \bibinfo {pages} {20200265} (\bibinfo
  {year} {2021})},\ \Eprint
  {http://arxiv.org/abs/https://royalsocietypublishing.org/doi/pdf/10.1098/rstb.2020.0265}
  {https://royalsocietypublishing.org/doi/pdf/10.1098/rstb.2020.0265}
  \BibitemShut {NoStop}%
\bibitem [{\citenamefont {Cao}\ \emph {et~al.}(2020)\citenamefont {Cao},
  \citenamefont {Zhang}, \citenamefont {Lu}, \citenamefont {Pfeiffer},
  \citenamefont {Jia}, \citenamefont {Song},\ and\ \citenamefont
  {Zeng}}]{Ca20}%
  \BibitemOpen
  \bibfield  {author} {\bibinfo {author} {\bibfnamefont {Z.}~\bibnamefont
  {Cao}}, \bibinfo {author} {\bibfnamefont {Q.}~\bibnamefont {Zhang}}, \bibinfo
  {author} {\bibfnamefont {X.}~\bibnamefont {Lu}}, \bibinfo {author}
  {\bibfnamefont {D.}~\bibnamefont {Pfeiffer}}, \bibinfo {author}
  {\bibfnamefont {Z.}~\bibnamefont {Jia}}, \bibinfo {author} {\bibfnamefont
  {H.}~\bibnamefont {Song}}, \ and\ \bibinfo {author} {\bibfnamefont {D.~D.}\
  \bibnamefont {Zeng}},\ }\href {\doibase 10.1101/2020.01.27.20018952}
  {\bibfield  {journal} {\bibinfo  {journal} {medRxiv}\ } (\bibinfo {year}
  {2020}),\ 10.1101/2020.01.27.20018952},\ \Eprint
  {http://arxiv.org/abs/https://www.medrxiv.org/content/early/2020/01/29/2020.01.27.20018952.full.pdf}
  {https://www.medrxiv.org/content/early/2020/01/29/2020.01.27.20018952.full.pdf}
  \BibitemShut {NoStop}%
\bibitem [{\citenamefont {Tang}\ \emph {et~al.}(2020)\citenamefont {Tang},
  \citenamefont {Wang}, \citenamefont {Li}, \citenamefont {Bragazzi},
  \citenamefont {Tang}, \citenamefont {Xiao},\ and\ \citenamefont {Wu}}]{Ta20}%
  \BibitemOpen
  \bibfield  {author} {\bibinfo {author} {\bibfnamefont {B.}~\bibnamefont
  {Tang}}, \bibinfo {author} {\bibfnamefont {X.}~\bibnamefont {Wang}}, \bibinfo
  {author} {\bibfnamefont {Q.}~\bibnamefont {Li}}, \bibinfo {author}
  {\bibfnamefont {N.~L.}\ \bibnamefont {Bragazzi}}, \bibinfo {author}
  {\bibfnamefont {S.}~\bibnamefont {Tang}}, \bibinfo {author} {\bibfnamefont
  {Y.}~\bibnamefont {Xiao}}, \ and\ \bibinfo {author} {\bibfnamefont
  {J.}~\bibnamefont {Wu}},\ }\href {\doibase 10.3390/jcm9020462} {\bibfield
  {journal} {\bibinfo  {journal} {Journal of clinical medicine}\ }\textbf
  {\bibinfo {volume} {9}},\ \bibinfo {pages} {462} (\bibinfo {year} {2020})},\
  \bibinfo {note} {32046137[pmid]}\BibitemShut {NoStop}%
\bibitem [{\citenamefont {Danon}\ \emph {et~al.}(2021)\citenamefont {Danon},
  \citenamefont {Brooks-Pollock}, \citenamefont {Bailey},\ and\ \citenamefont
  {Keeling}}]{Da21}%
  \BibitemOpen
  \bibfield  {author} {\bibinfo {author} {\bibfnamefont {L.}~\bibnamefont
  {Danon}}, \bibinfo {author} {\bibfnamefont {E.}~\bibnamefont
  {Brooks-Pollock}}, \bibinfo {author} {\bibfnamefont {M.}~\bibnamefont
  {Bailey}}, \ and\ \bibinfo {author} {\bibfnamefont {M.}~\bibnamefont
  {Keeling}},\ }\href {\doibase 10.1098/rstb.2020.0272} {\bibfield  {journal}
  {\bibinfo  {journal} {Philosophical Transactions of the Royal Society B:
  Biological Sciences}\ }\textbf {\bibinfo {volume} {376}},\ \bibinfo {pages}
  {20200272} (\bibinfo {year} {2021})},\ \Eprint
  {http://arxiv.org/abs/https://royalsocietypublishing.org/doi/pdf/10.1098/rstb.2020.0272}
  {https://royalsocietypublishing.org/doi/pdf/10.1098/rstb.2020.0272}
  \BibitemShut {NoStop}%
\bibitem [{\citenamefont {Wu}\ \emph {et~al.}(2020)\citenamefont {Wu},
  \citenamefont {Leung},\ and\ \citenamefont {Leung}}]{Wu20}%
  \BibitemOpen
  \bibfield  {author} {\bibinfo {author} {\bibfnamefont {J.~T.}\ \bibnamefont
  {Wu}}, \bibinfo {author} {\bibfnamefont {K.}~\bibnamefont {Leung}}, \ and\
  \bibinfo {author} {\bibfnamefont {G.~M.}\ \bibnamefont {Leung}},\ }\href
  {\doibase 10.1016/S0140-6736(20)30260-9} {\bibfield  {journal} {\bibinfo
  {journal} {The Lancet}\ }\textbf {\bibinfo {volume} {395}},\ \bibinfo {pages}
  {689} (\bibinfo {year} {2020})}\BibitemShut {NoStop}%
\bibitem [{\citenamefont {Riou}\ and\ \citenamefont {Althaus}(2020)}]{Ri20}%
  \BibitemOpen
  \bibfield  {author} {\bibinfo {author} {\bibfnamefont {J.}~\bibnamefont
  {Riou}}\ and\ \bibinfo {author} {\bibfnamefont {C.~L.}\ \bibnamefont
  {Althaus}},\ }\href {\doibase 10.2807/1560-7917.ES.2020.25.4.2000058}
  {\bibfield  {journal} {\bibinfo  {journal} {Euro surveillance : bulletin
  Europeen sur les maladies transmissibles = European communicable disease
  bulletin}\ }\textbf {\bibinfo {volume} {25}},\ \bibinfo {pages} {2000058}
  (\bibinfo {year} {2020})}\BibitemShut {NoStop}%
\bibitem [{\citenamefont {Sanche}\ \emph {et~al.}(2020)\citenamefont {Sanche},
  \citenamefont {Lin}, \citenamefont {Xu}, \citenamefont {Romero-Severson},
  \citenamefont {Hengartner},\ and\ \citenamefont {Ke}}]{Sa20}%
  \BibitemOpen
  \bibfield  {author} {\bibinfo {author} {\bibfnamefont {S.}~\bibnamefont
  {Sanche}}, \bibinfo {author} {\bibfnamefont {Y.~T.}\ \bibnamefont {Lin}},
  \bibinfo {author} {\bibfnamefont {C.}~\bibnamefont {Xu}}, \bibinfo {author}
  {\bibfnamefont {E.}~\bibnamefont {Romero-Severson}}, \bibinfo {author}
  {\bibfnamefont {N.}~\bibnamefont {Hengartner}}, \ and\ \bibinfo {author}
  {\bibfnamefont {R.}~\bibnamefont {Ke}},\ }\href {\doibase
  10.3201/eid2607.200282} {\bibfield  {journal} {\bibinfo  {journal} {Emerging
  infectious diseases}\ }\textbf {\bibinfo {volume} {26}},\ \bibinfo {pages}
  {1470—1477} (\bibinfo {year} {2020})}\BibitemShut {NoStop}%
\bibitem [{\citenamefont {Chinazzi}\ \emph {et~al.}(2020)\citenamefont
  {Chinazzi}, \citenamefont {Davis}, \citenamefont {Ajelli}, \citenamefont
  {Gioannini}, \citenamefont {Litvinova}, \citenamefont {Merler}, \citenamefont
  {Pastore~y Piontti}, \citenamefont {Mu}, \citenamefont {Rossi}, \citenamefont
  {Sun}, \citenamefont {Viboud}, \citenamefont {Xiong}, \citenamefont {Yu},
  \citenamefont {Halloran}, \citenamefont {Longini},\ and\ \citenamefont
  {Vespignani}}]{Ch20}%
  \BibitemOpen
  \bibfield  {author} {\bibinfo {author} {\bibfnamefont {M.}~\bibnamefont
  {Chinazzi}}, \bibinfo {author} {\bibfnamefont {J.~T.}\ \bibnamefont {Davis}},
  \bibinfo {author} {\bibfnamefont {M.}~\bibnamefont {Ajelli}}, \bibinfo
  {author} {\bibfnamefont {C.}~\bibnamefont {Gioannini}}, \bibinfo {author}
  {\bibfnamefont {M.}~\bibnamefont {Litvinova}}, \bibinfo {author}
  {\bibfnamefont {S.}~\bibnamefont {Merler}}, \bibinfo {author} {\bibfnamefont
  {A.}~\bibnamefont {Pastore~y Piontti}}, \bibinfo {author} {\bibfnamefont
  {K.}~\bibnamefont {Mu}}, \bibinfo {author} {\bibfnamefont {L.}~\bibnamefont
  {Rossi}}, \bibinfo {author} {\bibfnamefont {K.}~\bibnamefont {Sun}}, \bibinfo
  {author} {\bibfnamefont {C.}~\bibnamefont {Viboud}}, \bibinfo {author}
  {\bibfnamefont {X.}~\bibnamefont {Xiong}}, \bibinfo {author} {\bibfnamefont
  {H.}~\bibnamefont {Yu}}, \bibinfo {author} {\bibfnamefont {M.~E.}\
  \bibnamefont {Halloran}}, \bibinfo {author} {\bibfnamefont {I.~M.}\
  \bibnamefont {Longini}}, \ and\ \bibinfo {author} {\bibfnamefont
  {A.}~\bibnamefont {Vespignani}},\ }\href {\doibase 10.1126/science.aba9757}
  {\bibfield  {journal} {\bibinfo  {journal} {Science}\ }\textbf {\bibinfo
  {volume} {368}},\ \bibinfo {pages} {395} (\bibinfo {year} {2020})},\ \Eprint
  {http://arxiv.org/abs/https://science.sciencemag.org/content/368/6489/395.full.pdf}
  {https://science.sciencemag.org/content/368/6489/395.full.pdf} \BibitemShut
  {NoStop}%
\bibitem [{\citenamefont {Wong}\ \emph {et~al.}(2020)\citenamefont {Wong},
  \citenamefont {Weiner}, \citenamefont {Tkachenko}, \citenamefont {Elbanna},
  \citenamefont {Maslov},\ and\ \citenamefont {Goldenfeld}}]{Wo20}%
  \BibitemOpen
  \bibfield  {author} {\bibinfo {author} {\bibfnamefont {G.~N.}\ \bibnamefont
  {Wong}}, \bibinfo {author} {\bibfnamefont {Z.~J.}\ \bibnamefont {Weiner}},
  \bibinfo {author} {\bibfnamefont {A.~V.}\ \bibnamefont {Tkachenko}}, \bibinfo
  {author} {\bibfnamefont {A.}~\bibnamefont {Elbanna}}, \bibinfo {author}
  {\bibfnamefont {S.}~\bibnamefont {Maslov}}, \ and\ \bibinfo {author}
  {\bibfnamefont {N.}~\bibnamefont {Goldenfeld}},\ }\href {\doibase
  10.1103/PhysRevX.10.041033} {\bibfield  {journal} {\bibinfo  {journal} {Phys.
  Rev. X}\ }\textbf {\bibinfo {volume} {10}},\ \bibinfo {pages} {041033}
  (\bibinfo {year} {2020})}\BibitemShut {NoStop}%
\bibitem [{\citenamefont {Chang}\ \emph {et~al.}(2021)\citenamefont {Chang},
  \citenamefont {Pierson}, \citenamefont {Koh}, \citenamefont {Gerardin},
  \citenamefont {Redbird}, \citenamefont {Grusky},\ and\ \citenamefont
  {Leskovec}}]{Ch21}%
  \BibitemOpen
  \bibfield  {author} {\bibinfo {author} {\bibfnamefont {S.}~\bibnamefont
  {Chang}}, \bibinfo {author} {\bibfnamefont {E.}~\bibnamefont {Pierson}},
  \bibinfo {author} {\bibfnamefont {P.~W.}\ \bibnamefont {Koh}}, \bibinfo
  {author} {\bibfnamefont {J.}~\bibnamefont {Gerardin}}, \bibinfo {author}
  {\bibfnamefont {B.}~\bibnamefont {Redbird}}, \bibinfo {author} {\bibfnamefont
  {D.}~\bibnamefont {Grusky}}, \ and\ \bibinfo {author} {\bibfnamefont
  {J.}~\bibnamefont {Leskovec}},\ }\href {\doibase 10.1038/s41586-020-2923-3}
  {\bibfield  {journal} {\bibinfo  {journal} {Nature}\ }\textbf {\bibinfo
  {volume} {589}},\ \bibinfo {pages} {82} (\bibinfo {year} {2021})}\BibitemShut
  {NoStop}%
\bibitem [{\citenamefont {Borremans}\ \emph {et~al.}(2017)\citenamefont
  {Borremans}, \citenamefont {Reijniers}, \citenamefont {Hens},\ and\
  \citenamefont {Leirs}}]{Borr17}%
  \BibitemOpen
  \bibfield  {author} {\bibinfo {author} {\bibfnamefont {B.}~\bibnamefont
  {Borremans}}, \bibinfo {author} {\bibfnamefont {J.}~\bibnamefont
  {Reijniers}}, \bibinfo {author} {\bibfnamefont {N.}~\bibnamefont {Hens}}, \
  and\ \bibinfo {author} {\bibfnamefont {H.}~\bibnamefont {Leirs}},\ }\href
  {\doibase 10.1098/rsos.171308} {\bibfield  {journal} {\bibinfo  {journal}
  {Royal Society Open Science}\ }\textbf {\bibinfo {volume} {4}},\ \bibinfo
  {pages} {171308} (\bibinfo {year} {2017})},\ \Eprint
  {http://arxiv.org/abs/https://royalsocietypublishing.org/doi/pdf/10.1098/rsos.171308}
  {https://royalsocietypublishing.org/doi/pdf/10.1098/rsos.171308} \BibitemShut
  {NoStop}%
\bibitem [{\citenamefont {Fuk{\'{s}}}\ \emph {et~al.}(2006)\citenamefont
  {Fuk{\'{s}}}, \citenamefont {Lawniczak},\ and\ \citenamefont
  {Duchesne}}]{Fu06}%
  \BibitemOpen
  \bibfield  {author} {\bibinfo {author} {\bibfnamefont {H.}~\bibnamefont
  {Fuk{\'{s}}}}, \bibinfo {author} {\bibfnamefont {A.~T.}\ \bibnamefont
  {Lawniczak}}, \ and\ \bibinfo {author} {\bibfnamefont {R.}~\bibnamefont
  {Duchesne}},\ }\href {\doibase 10.1140/epjb/e2006-00136-7} {\bibfield
  {journal} {\bibinfo  {journal} {The European Physical Journal B - Condensed
  Matter and Complex Systems}\ }\textbf {\bibinfo {volume} {50}},\ \bibinfo
  {pages} {209} (\bibinfo {year} {2006})}\BibitemShut {NoStop}%
\bibitem [{\citenamefont {{World Health
  Organisation}}(2021{\natexlab{b}})}]{WHO2}%
  \BibitemOpen
  \bibfield  {author} {\bibinfo {author} {\bibnamefont {{World Health
  Organisation}}},\ }\href@noop {} {} (\bibinfo {year} {2021}{\natexlab{b}}),\
  \bibinfo {note}
  {\url{https://www.who.int/emergencies/diseases/novel-coronavirus-2019/technical-guidance/maintaining-essential-health-services-and-systems},
  Last accessed on 2021-09-09}\BibitemShut {NoStop}%
\bibitem [{\citenamefont {Gilbert}\ \emph {et~al.}(2020)\citenamefont
  {Gilbert}, \citenamefont {Pullano}, \citenamefont {Pinotti}, \citenamefont
  {Valdano}, \citenamefont {Poletto}, \citenamefont {Bo{\"e}lle}, \citenamefont
  {D'Ortenzio}, \citenamefont {Yazdanpanah}, \citenamefont {Eholie},
  \citenamefont {Altmann}, \citenamefont {Gutierrez}, \citenamefont {Kraemer},\
  and\ \citenamefont {Colizza}}]{Gi20}%
  \BibitemOpen
  \bibfield  {author} {\bibinfo {author} {\bibfnamefont {M.}~\bibnamefont
  {Gilbert}}, \bibinfo {author} {\bibfnamefont {G.}~\bibnamefont {Pullano}},
  \bibinfo {author} {\bibfnamefont {F.}~\bibnamefont {Pinotti}}, \bibinfo
  {author} {\bibfnamefont {E.}~\bibnamefont {Valdano}}, \bibinfo {author}
  {\bibfnamefont {C.}~\bibnamefont {Poletto}}, \bibinfo {author} {\bibfnamefont
  {P.-Y.}\ \bibnamefont {Bo{\"e}lle}}, \bibinfo {author} {\bibfnamefont
  {E.}~\bibnamefont {D'Ortenzio}}, \bibinfo {author} {\bibfnamefont
  {Y.}~\bibnamefont {Yazdanpanah}}, \bibinfo {author} {\bibfnamefont {S.~P.}\
  \bibnamefont {Eholie}}, \bibinfo {author} {\bibfnamefont {M.}~\bibnamefont
  {Altmann}}, \bibinfo {author} {\bibfnamefont {B.}~\bibnamefont {Gutierrez}},
  \bibinfo {author} {\bibfnamefont {M.~U.~G.}\ \bibnamefont {Kraemer}}, \ and\
  \bibinfo {author} {\bibfnamefont {V.}~\bibnamefont {Colizza}},\ }\href
  {\doibase 10.1016/S0140-6736(20)30411-6} {\bibfield  {journal} {\bibinfo
  {journal} {The Lancet}\ }\textbf {\bibinfo {volume} {395}},\ \bibinfo {pages}
  {871} (\bibinfo {year} {2020})}\BibitemShut {NoStop}%
\bibitem [{\citenamefont {Pullano}\ \emph {et~al.}(2020)\citenamefont
  {Pullano}, \citenamefont {Pinotti}, \citenamefont {Valdano}, \citenamefont
  {Bo{\"e}lle}, \citenamefont {Poletto},\ and\ \citenamefont {Colizza}}]{Pu20}%
  \BibitemOpen
  \bibfield  {author} {\bibinfo {author} {\bibfnamefont {G.}~\bibnamefont
  {Pullano}}, \bibinfo {author} {\bibfnamefont {F.}~\bibnamefont {Pinotti}},
  \bibinfo {author} {\bibfnamefont {E.}~\bibnamefont {Valdano}}, \bibinfo
  {author} {\bibfnamefont {P.-Y.}\ \bibnamefont {Bo{\"e}lle}}, \bibinfo
  {author} {\bibfnamefont {C.}~\bibnamefont {Poletto}}, \ and\ \bibinfo
  {author} {\bibfnamefont {V.}~\bibnamefont {Colizza}},\ }\href {\doibase
  10.2807/1560-7917.ES.2020.25.4.2000057} {\bibfield  {journal} {\bibinfo
  {journal} {Euro surveillance : bulletin Europeen sur les maladies
  transmissibles = European communicable disease bulletin}\ }\textbf {\bibinfo
  {volume} {25}},\ \bibinfo {pages} {2000057} (\bibinfo {year} {2020})},\
  \bibinfo {note} {32019667[pmid]}\BibitemShut {NoStop}%
\bibitem [{\citenamefont {Balcan}\ \emph {et~al.}(2009)\citenamefont {Balcan},
  \citenamefont {Colizza}, \citenamefont {Gon{\c c}alves}, \citenamefont {Hu},
  \citenamefont {Ramasco},\ and\ \citenamefont {Vespignani}}]{Ba09}%
  \BibitemOpen
  \bibfield  {author} {\bibinfo {author} {\bibfnamefont {D.}~\bibnamefont
  {Balcan}}, \bibinfo {author} {\bibfnamefont {V.}~\bibnamefont {Colizza}},
  \bibinfo {author} {\bibfnamefont {B.}~\bibnamefont {Gon{\c c}alves}},
  \bibinfo {author} {\bibfnamefont {H.}~\bibnamefont {Hu}}, \bibinfo {author}
  {\bibfnamefont {J.~J.}\ \bibnamefont {Ramasco}}, \ and\ \bibinfo {author}
  {\bibfnamefont {A.}~\bibnamefont {Vespignani}},\ }\href {\doibase
  10.1073/pnas.0906910106} {\bibfield  {journal} {\bibinfo  {journal}
  {Proceedings of the National Academy of Sciences}\ }\textbf {\bibinfo
  {volume} {106}},\ \bibinfo {pages} {21484} (\bibinfo {year} {2009})},\
  \Eprint
  {http://arxiv.org/abs/https://www.pnas.org/content/106/51/21484.full.pdf}
  {https://www.pnas.org/content/106/51/21484.full.pdf} \BibitemShut {NoStop}%
\bibitem [{\citenamefont {Danon}\ \emph {et~al.}(2009)\citenamefont {Danon},
  \citenamefont {House},\ and\ \citenamefont {Keeling}}]{Da09}%
  \BibitemOpen
  \bibfield  {author} {\bibinfo {author} {\bibfnamefont {L.}~\bibnamefont
  {Danon}}, \bibinfo {author} {\bibfnamefont {T.}~\bibnamefont {House}}, \ and\
  \bibinfo {author} {\bibfnamefont {M.~J.}\ \bibnamefont {Keeling}},\ }\href
  {\doibase https://doi.org/10.1016/j.epidem.2009.11.002} {\bibfield  {journal}
  {\bibinfo  {journal} {Epidemics}\ }\textbf {\bibinfo {volume} {1}},\ \bibinfo
  {pages} {250} (\bibinfo {year} {2009})}\BibitemShut {NoStop}%
\bibitem [{\citenamefont {Dordević}\ \emph {et~al.}(2021)\citenamefont
  {Dordević}, \citenamefont {Papić},\ and\ \citenamefont {Šuvak}}]{Dor21}%
  \BibitemOpen
  \bibfield  {author} {\bibinfo {author} {\bibfnamefont {J.}~\bibnamefont
  {Dordević}}, \bibinfo {author} {\bibfnamefont {I.}~\bibnamefont {Papić}}, \
  and\ \bibinfo {author} {\bibfnamefont {N.}~\bibnamefont {Šuvak}},\ }\href
  {\doibase https://doi.org/10.1016/j.chaos.2021.110991} {\bibfield  {journal}
  {\bibinfo  {journal} {Chaos, Solitons and Fractals}\ }\textbf {\bibinfo
  {volume} {148}},\ \bibinfo {pages} {110991} (\bibinfo {year}
  {2021})}\BibitemShut {NoStop}%
\bibitem [{\citenamefont {Adak}\ \emph {et~al.}(2021)\citenamefont {Adak},
  \citenamefont {Majumder},\ and\ \citenamefont {Bairagi}}]{Deb21}%
  \BibitemOpen
  \bibfield  {author} {\bibinfo {author} {\bibfnamefont {D.}~\bibnamefont
  {Adak}}, \bibinfo {author} {\bibfnamefont {A.}~\bibnamefont {Majumder}}, \
  and\ \bibinfo {author} {\bibfnamefont {N.}~\bibnamefont {Bairagi}},\ }\href
  {\doibase https://doi.org/10.1016/j.chaos.2020.110381} {\bibfield  {journal}
  {\bibinfo  {journal} {Chaos, Solitons and Fractals}\ }\textbf {\bibinfo
  {volume} {142}},\ \bibinfo {pages} {110381} (\bibinfo {year}
  {2021})}\BibitemShut {NoStop}%
\bibitem [{\citenamefont {Engbert}\ \emph {et~al.}(2020)\citenamefont
  {Engbert}, \citenamefont {Rabe}, \citenamefont {Kliegl},\ and\ \citenamefont
  {Reich}}]{Eng20}%
  \BibitemOpen
  \bibfield  {author} {\bibinfo {author} {\bibfnamefont {R.}~\bibnamefont
  {Engbert}}, \bibinfo {author} {\bibfnamefont {M.~M.}\ \bibnamefont {Rabe}},
  \bibinfo {author} {\bibfnamefont {R.}~\bibnamefont {Kliegl}}, \ and\ \bibinfo
  {author} {\bibfnamefont {S.}~\bibnamefont {Reich}},\ }\href {\doibase
  10.1007/s11538-020-00834-8} {\bibfield  {journal} {\bibinfo  {journal}
  {Bulletin of Mathematical Biology}\ }\textbf {\bibinfo {volume} {83}},\
  \bibinfo {pages} {1} (\bibinfo {year} {2020})}\BibitemShut {NoStop}%
\bibitem [{\citenamefont {Hazarie}\ \emph {et~al.}(2021)\citenamefont
  {Hazarie}, \citenamefont {Soriano-Pa{\~{n}}os}, \citenamefont {Arenas},
  \citenamefont {G{\'o}mez-Garde{\~{n}}es},\ and\ \citenamefont
  {Ghoshal}}]{Ha21}%
  \BibitemOpen
  \bibfield  {author} {\bibinfo {author} {\bibfnamefont {S.}~\bibnamefont
  {Hazarie}}, \bibinfo {author} {\bibfnamefont {D.}~\bibnamefont
  {Soriano-Pa{\~{n}}os}}, \bibinfo {author} {\bibfnamefont {A.}~\bibnamefont
  {Arenas}}, \bibinfo {author} {\bibfnamefont {J.}~\bibnamefont
  {G{\'o}mez-Garde{\~{n}}es}}, \ and\ \bibinfo {author} {\bibfnamefont
  {G.}~\bibnamefont {Ghoshal}},\ }\href {\doibase 10.1038/s42005-021-00679-0}
  {\bibfield  {journal} {\bibinfo  {journal} {Communications Physics}\ }\textbf
  {\bibinfo {volume} {4}},\ \bibinfo {pages} {191} (\bibinfo {year}
  {2021})}\BibitemShut {NoStop}%
\bibitem [{\citenamefont {Li}\ \emph {et~al.}(2020)\citenamefont {Li},
  \citenamefont {Guan}, \citenamefont {Wu}, \citenamefont {Wang}, \citenamefont
  {Zhou}, \citenamefont {Tong}, \citenamefont {Ren}, \citenamefont {Leung},
  \citenamefont {Lau}, \citenamefont {Wong}, \citenamefont {Xing},
  \citenamefont {Xiang}, \citenamefont {Wu}, \citenamefont {Li}, \citenamefont
  {Chen}, \citenamefont {Li}, \citenamefont {Liu}, \citenamefont {Zhao},
  \citenamefont {Liu}, \citenamefont {Tu}, \citenamefont {Chen}, \citenamefont
  {Jin}, \citenamefont {Yang}, \citenamefont {Wang}, \citenamefont {Zhou},
  \citenamefont {Wang}, \citenamefont {Liu}, \citenamefont {Luo}, \citenamefont
  {Liu}, \citenamefont {Shao}, \citenamefont {Li}, \citenamefont {Tao},
  \citenamefont {Yang}, \citenamefont {Deng}, \citenamefont {Liu},
  \citenamefont {Ma}, \citenamefont {Zhang}, \citenamefont {Shi}, \citenamefont
  {Lam}, \citenamefont {Wu}, \citenamefont {Gao}, \citenamefont {Cowling},
  \citenamefont {Yang}, \citenamefont {Leung},\ and\ \citenamefont
  {Feng}}]{Li20}%
  \BibitemOpen
  \bibfield  {author} {\bibinfo {author} {\bibfnamefont {Q.}~\bibnamefont
  {Li}}, \bibinfo {author} {\bibfnamefont {X.}~\bibnamefont {Guan}}, \bibinfo
  {author} {\bibfnamefont {P.}~\bibnamefont {Wu}}, \bibinfo {author}
  {\bibfnamefont {X.}~\bibnamefont {Wang}}, \bibinfo {author} {\bibfnamefont
  {L.}~\bibnamefont {Zhou}}, \bibinfo {author} {\bibfnamefont {Y.}~\bibnamefont
  {Tong}}, \bibinfo {author} {\bibfnamefont {R.}~\bibnamefont {Ren}}, \bibinfo
  {author} {\bibfnamefont {K.~S.}\ \bibnamefont {Leung}}, \bibinfo {author}
  {\bibfnamefont {E.~H.}\ \bibnamefont {Lau}}, \bibinfo {author} {\bibfnamefont
  {J.~Y.}\ \bibnamefont {Wong}}, \bibinfo {author} {\bibfnamefont
  {X.}~\bibnamefont {Xing}}, \bibinfo {author} {\bibfnamefont {N.}~\bibnamefont
  {Xiang}}, \bibinfo {author} {\bibfnamefont {Y.}~\bibnamefont {Wu}}, \bibinfo
  {author} {\bibfnamefont {C.}~\bibnamefont {Li}}, \bibinfo {author}
  {\bibfnamefont {Q.}~\bibnamefont {Chen}}, \bibinfo {author} {\bibfnamefont
  {D.}~\bibnamefont {Li}}, \bibinfo {author} {\bibfnamefont {T.}~\bibnamefont
  {Liu}}, \bibinfo {author} {\bibfnamefont {J.}~\bibnamefont {Zhao}}, \bibinfo
  {author} {\bibfnamefont {M.}~\bibnamefont {Liu}}, \bibinfo {author}
  {\bibfnamefont {W.}~\bibnamefont {Tu}}, \bibinfo {author} {\bibfnamefont
  {C.}~\bibnamefont {Chen}}, \bibinfo {author} {\bibfnamefont {L.}~\bibnamefont
  {Jin}}, \bibinfo {author} {\bibfnamefont {R.}~\bibnamefont {Yang}}, \bibinfo
  {author} {\bibfnamefont {Q.}~\bibnamefont {Wang}}, \bibinfo {author}
  {\bibfnamefont {S.}~\bibnamefont {Zhou}}, \bibinfo {author} {\bibfnamefont
  {R.}~\bibnamefont {Wang}}, \bibinfo {author} {\bibfnamefont {H.}~\bibnamefont
  {Liu}}, \bibinfo {author} {\bibfnamefont {Y.}~\bibnamefont {Luo}}, \bibinfo
  {author} {\bibfnamefont {Y.}~\bibnamefont {Liu}}, \bibinfo {author}
  {\bibfnamefont {G.}~\bibnamefont {Shao}}, \bibinfo {author} {\bibfnamefont
  {H.}~\bibnamefont {Li}}, \bibinfo {author} {\bibfnamefont {Z.}~\bibnamefont
  {Tao}}, \bibinfo {author} {\bibfnamefont {Y.}~\bibnamefont {Yang}}, \bibinfo
  {author} {\bibfnamefont {Z.}~\bibnamefont {Deng}}, \bibinfo {author}
  {\bibfnamefont {B.}~\bibnamefont {Liu}}, \bibinfo {author} {\bibfnamefont
  {Z.}~\bibnamefont {Ma}}, \bibinfo {author} {\bibfnamefont {Y.}~\bibnamefont
  {Zhang}}, \bibinfo {author} {\bibfnamefont {G.}~\bibnamefont {Shi}}, \bibinfo
  {author} {\bibfnamefont {T.~T.}\ \bibnamefont {Lam}}, \bibinfo {author}
  {\bibfnamefont {J.~T.}\ \bibnamefont {Wu}}, \bibinfo {author} {\bibfnamefont
  {G.~F.}\ \bibnamefont {Gao}}, \bibinfo {author} {\bibfnamefont {B.~J.}\
  \bibnamefont {Cowling}}, \bibinfo {author} {\bibfnamefont {B.}~\bibnamefont
  {Yang}}, \bibinfo {author} {\bibfnamefont {G.~M.}\ \bibnamefont {Leung}}, \
  and\ \bibinfo {author} {\bibfnamefont {Z.}~\bibnamefont {Feng}},\ }\href
  {\doibase 10.1056/NEJMoa2001316} {\bibfield  {journal} {\bibinfo  {journal}
  {New England Journal of Medicine}\ }\textbf {\bibinfo {volume} {382}},\
  \bibinfo {pages} {1199} (\bibinfo {year} {2020})},\ \bibinfo {note} {pMID:
  31995857},\ \Eprint
  {http://arxiv.org/abs/https://doi.org/10.1056/NEJMoa2001316}
  {https://doi.org/10.1056/NEJMoa2001316} \BibitemShut {NoStop}%
\bibitem [{\citenamefont {Guan}\ \emph {et~al.}(2020)\citenamefont {Guan},
  \citenamefont {Ni}, \citenamefont {Hu}, \citenamefont {Liang}, \citenamefont
  {Ou}, \citenamefont {He}, \citenamefont {Liu}, \citenamefont {Shan},
  \citenamefont {Lei}, \citenamefont {Hui}, \citenamefont {Du}, \citenamefont
  {Li}, \citenamefont {Zeng}, \citenamefont {Yuen}, \citenamefont {Chen},
  \citenamefont {Tang}, \citenamefont {Wang}, \citenamefont {Chen},
  \citenamefont {Xiang}, \citenamefont {Li}, \citenamefont {Wang},
  \citenamefont {Liang}, \citenamefont {Peng}, \citenamefont {Wei},
  \citenamefont {Liu}, \citenamefont {Hu}, \citenamefont {Peng}, \citenamefont
  {Wang}, \citenamefont {Liu}, \citenamefont {Chen}, \citenamefont {Li},
  \citenamefont {Zheng}, \citenamefont {Qiu}, \citenamefont {Luo},
  \citenamefont {Ye}, \citenamefont {Zhu},\ and\ \citenamefont {Zhong}}]{Gu20}%
  \BibitemOpen
  \bibfield  {author} {\bibinfo {author} {\bibfnamefont {W.-j.}\ \bibnamefont
  {Guan}}, \bibinfo {author} {\bibfnamefont {Z.-y.}\ \bibnamefont {Ni}},
  \bibinfo {author} {\bibfnamefont {Y.}~\bibnamefont {Hu}}, \bibinfo {author}
  {\bibfnamefont {W.-h.}\ \bibnamefont {Liang}}, \bibinfo {author}
  {\bibfnamefont {C.-q.}\ \bibnamefont {Ou}}, \bibinfo {author} {\bibfnamefont
  {J.-x.}\ \bibnamefont {He}}, \bibinfo {author} {\bibfnamefont
  {L.}~\bibnamefont {Liu}}, \bibinfo {author} {\bibfnamefont {H.}~\bibnamefont
  {Shan}}, \bibinfo {author} {\bibfnamefont {C.-l.}\ \bibnamefont {Lei}},
  \bibinfo {author} {\bibfnamefont {D.~S.}\ \bibnamefont {Hui}}, \bibinfo
  {author} {\bibfnamefont {B.}~\bibnamefont {Du}}, \bibinfo {author}
  {\bibfnamefont {L.-j.}\ \bibnamefont {Li}}, \bibinfo {author} {\bibfnamefont
  {G.}~\bibnamefont {Zeng}}, \bibinfo {author} {\bibfnamefont {K.-Y.}\
  \bibnamefont {Yuen}}, \bibinfo {author} {\bibfnamefont {R.-c.}\ \bibnamefont
  {Chen}}, \bibinfo {author} {\bibfnamefont {C.-l.}\ \bibnamefont {Tang}},
  \bibinfo {author} {\bibfnamefont {T.}~\bibnamefont {Wang}}, \bibinfo {author}
  {\bibfnamefont {P.-y.}\ \bibnamefont {Chen}}, \bibinfo {author}
  {\bibfnamefont {J.}~\bibnamefont {Xiang}}, \bibinfo {author} {\bibfnamefont
  {S.-y.}\ \bibnamefont {Li}}, \bibinfo {author} {\bibfnamefont {J.-l.}\
  \bibnamefont {Wang}}, \bibinfo {author} {\bibfnamefont {Z.-j.}\ \bibnamefont
  {Liang}}, \bibinfo {author} {\bibfnamefont {Y.-x.}\ \bibnamefont {Peng}},
  \bibinfo {author} {\bibfnamefont {L.}~\bibnamefont {Wei}}, \bibinfo {author}
  {\bibfnamefont {Y.}~\bibnamefont {Liu}}, \bibinfo {author} {\bibfnamefont
  {Y.-h.}\ \bibnamefont {Hu}}, \bibinfo {author} {\bibfnamefont
  {P.}~\bibnamefont {Peng}}, \bibinfo {author} {\bibfnamefont {J.-m.}\
  \bibnamefont {Wang}}, \bibinfo {author} {\bibfnamefont {J.-y.}\ \bibnamefont
  {Liu}}, \bibinfo {author} {\bibfnamefont {Z.}~\bibnamefont {Chen}}, \bibinfo
  {author} {\bibfnamefont {G.}~\bibnamefont {Li}}, \bibinfo {author}
  {\bibfnamefont {Z.-j.}\ \bibnamefont {Zheng}}, \bibinfo {author}
  {\bibfnamefont {S.-q.}\ \bibnamefont {Qiu}}, \bibinfo {author} {\bibfnamefont
  {J.}~\bibnamefont {Luo}}, \bibinfo {author} {\bibfnamefont {C.-j.}\
  \bibnamefont {Ye}}, \bibinfo {author} {\bibfnamefont {S.-y.}\ \bibnamefont
  {Zhu}}, \ and\ \bibinfo {author} {\bibfnamefont {N.-s.}\ \bibnamefont
  {Zhong}},\ }\href {\doibase 10.1056/NEJMoa2002032} {\bibfield  {journal}
  {\bibinfo  {journal} {New England Journal of Medicine}\ }\textbf {\bibinfo
  {volume} {382}},\ \bibinfo {pages} {1708} (\bibinfo {year} {2020})},\ \Eprint
  {http://arxiv.org/abs/https://doi.org/10.1056/NEJMoa2002032}
  {https://doi.org/10.1056/NEJMoa2002032} \BibitemShut {NoStop}%
\bibitem [{\citenamefont {Backer}\ \emph {et~al.}(2020)\citenamefont {Backer},
  \citenamefont {Klinkenberg},\ and\ \citenamefont {Wallinga}}]{Ba20}%
  \BibitemOpen
  \bibfield  {author} {\bibinfo {author} {\bibfnamefont {J.~A.}\ \bibnamefont
  {Backer}}, \bibinfo {author} {\bibfnamefont {D.}~\bibnamefont {Klinkenberg}},
  \ and\ \bibinfo {author} {\bibfnamefont {J.}~\bibnamefont {Wallinga}},\
  }\href {\doibase 10.2807/1560-7917.ES.2020.25.5.2000062} {\bibfield
  {journal} {\bibinfo  {journal} {Euro surveillance : bulletin Europeen sur les
  maladies transmissibles = European communicable disease bulletin}\ }\textbf
  {\bibinfo {volume} {25}},\ \bibinfo {pages} {2000062} (\bibinfo {year}
  {2020})},\ \bibinfo {note} {32046819[pmid]}\BibitemShut {NoStop}%
\bibitem [{\citenamefont {Tindale}\ \emph {et~al.}(2020)\citenamefont
  {Tindale}, \citenamefont {Stockdale}, \citenamefont {Coombe}, \citenamefont
  {Garlock}, \citenamefont {Lau}, \citenamefont {Saraswat}, \citenamefont
  {Zhang}, \citenamefont {Chen}, \citenamefont {Wallinga},\ and\ \citenamefont
  {Colijn}}]{Ti20}%
  \BibitemOpen
  \bibfield  {author} {\bibinfo {author} {\bibfnamefont {L.~C.}\ \bibnamefont
  {Tindale}}, \bibinfo {author} {\bibfnamefont {J.~E.}\ \bibnamefont
  {Stockdale}}, \bibinfo {author} {\bibfnamefont {M.}~\bibnamefont {Coombe}},
  \bibinfo {author} {\bibfnamefont {E.~S.}\ \bibnamefont {Garlock}}, \bibinfo
  {author} {\bibfnamefont {W.~Y.~V.}\ \bibnamefont {Lau}}, \bibinfo {author}
  {\bibfnamefont {M.}~\bibnamefont {Saraswat}}, \bibinfo {author}
  {\bibfnamefont {L.}~\bibnamefont {Zhang}}, \bibinfo {author} {\bibfnamefont
  {D.}~\bibnamefont {Chen}}, \bibinfo {author} {\bibfnamefont {J.}~\bibnamefont
  {Wallinga}}, \ and\ \bibinfo {author} {\bibfnamefont {C.}~\bibnamefont
  {Colijn}},\ }\href {\doibase 10.7554/eLife.57149} {\bibfield  {journal}
  {\bibinfo  {journal} {eLife}\ }\textbf {\bibinfo {volume} {9}},\ \bibinfo
  {pages} {e57149} (\bibinfo {year} {2020})},\ \bibinfo {note}
  {32568070[pmid]}\BibitemShut {NoStop}%
\bibitem [{\citenamefont {John}\ and\ \citenamefont {Samuel}(2000)}]{TJ00}%
  \BibitemOpen
  \bibfield  {author} {\bibinfo {author} {\bibfnamefont {T.~J.}\ \bibnamefont
  {John}}\ and\ \bibinfo {author} {\bibfnamefont {R.}~\bibnamefont {Samuel}},\
  }\href@noop {} {\bibfield  {journal} {\bibinfo  {journal} {Eur J Epidemiol}\
  }\textbf {\bibinfo {volume} {16}},\ \bibinfo {pages} {601} (\bibinfo {year}
  {2000})}\BibitemShut {NoStop}%
\bibitem [{\citenamefont {Rashid}\ \emph {et~al.}(2012)\citenamefont {Rashid},
  \citenamefont {Khandaker},\ and\ \citenamefont {Booy}}]{Rash12}%
  \BibitemOpen
  \bibfield  {author} {\bibinfo {author} {\bibfnamefont {H.}~\bibnamefont
  {Rashid}}, \bibinfo {author} {\bibfnamefont {G.}~\bibnamefont {Khandaker}}, \
  and\ \bibinfo {author} {\bibfnamefont {R.}~\bibnamefont {Booy}},\ }\href
  {https://journals.lww.com/co-infectiousdiseases/Fulltext/2012/06000/Vaccination_and_herd_immunity__what_more_do_we.3.aspx}
  {\bibfield  {journal} {\bibinfo  {journal} {Current Opinion in Infectious
  Diseases}\ }\textbf {\bibinfo {volume} {25}} (\bibinfo {year}
  {2012})}\BibitemShut {NoStop}%
\end{thebibliography}%

\end{document}